\documentclass[a4paper]{article}

\providecommand{\preprintno}[1]{}

\usepackage{feynmp}   %%% MetaPost

{\def\$#1: #2 ${#2}}
\def\Version/{1.5}
\def\Date/{May 1996}

\def\.#1{\texttt{#1}}
\def\WOPPER/{\.{WOPPER}}
\def\f77/{\.{FORTRAN-77}}
\def\hepawk/{\.{hepawk}}
\def\hepevt/{\.{/hepevt/}}
\def\HERWIG/{\.{HERWIG}}
\thicklines
\def\tablestrut{\vrule width 0pt height 12pt depth 3pt}

\newenvironment{example}[2]%
  {\def\examplecaption{#1}%
   \def\examplelabel{#2}%
   \begin{figure}%
   \begin{tabbing}%
   \tab\tab\tab\tab\tab\tab\tab\tab\tab\tab\kill}%
  {\end{tabbing}%
   \caption{\examplecaption}%
   \label{\examplelabel}%
   \end{figure}}
\def\tab{\quad\=}
\def\C{\`\it} % comments

\newenvironment{commands}%
 {\begin{list}{}%
   {\setlength{\leftmargin}{2em}%
    \setlength{\rightmargin}{2em}%
    \setlength{\itemindent}{-1em}%
    \setlength{\listparindent}{0pt}%
    }}%
 {\end{list}}
\newcommand{\meta}[1]{$\langle$\textit{#1}$\rangle$}

\begin{document}

\title{%
  \WOPPER/, Version \Version/: \\
  A Monte Carlo Event Generator for \\
  $e^+e^- \to (W^+W^-) \to 4f + n\gamma$ \\
  at LEP2 and beyond%
  \thanks{Supported by Bundesministerium f\"ur Forschung und
    Technologie, Germany}}%

\author{%
  Harald Anlauf%
    \thanks{Email: \texttt{anlauf@crunch.ikp.physik.th-darmstadt.de}},
  Panagiotis Manakos,
  Thorsten Ohl%
    \thanks{Email: \texttt{Thorsten.Ohl@Physik.TH-Darmstadt.de}}\\
  Technische Hochschule Darmstadt\\
  64289 Darmstadt, Germany\\
  \hfil\\
  Hans Dieter Dahmen\\
  Universit\"at Siegen\\
  57076 Siegen, Germany}

\preprintno{%
  IKDA 96/15\\
  hep-ph/9605457}
\date{May 1996}

%%%%%%%%%%%%%%%%%%%%%%%%%%%%%%%%%%%%%%%%%%%%%%%%%%%%%%%%%%%%%%%%%%%%%%%%

\maketitle
\begin{abstract}
  We describe the new version of the Monte Carlo event generator
  \WOPPER/ for four fermion production through $W$-pairs including
  resummed leading logarithmic QED radiative corrections.  Among the new
  features included are singly resonant background diagrams and
  anomalous triple gauge boson couplings.
\end{abstract}

%%%%%%%%%%%%%%%%%%%%%%%%%%%%%%%%%%%%%%%%%%%%%%%%%%%%%%%%%%%%%%%%%%%%
\section*{Program Summary:}

\begin{itemize}
\item{} \textbf{Title of program:} \WOPPER/, Version \Version/ (\Date/)
\item{} \textbf{Program obtainable from:}
  \texttt{crunch.ikp.physik.th-darmstadt.de} in the 
  directory \texttt{/pub/ohl/wopper} (using anonymous Internet ftp)
\item{} \textbf{Licensing provisions:} none
\item{} \textbf{Programming language used:} \f77/
\item{} \textbf{Computer/Operating System:}
  Any with a \f77/ environment
\item{} \textbf{Number of program lines in distributed program, including
    test data, etc.:} $\approx$ 10000 (Including comments)
\item{} \textbf{Keywords:} radiative corrections, $W$ pair production, $W$
  decays, multiphoton radiation, anomalous couplings.
\item{} \textbf{Nature of physical problem:}  Higher order leading
  logarithmic QED radiative corrections to $W$ pair production and decay
  at high energy $e^+e^-$ colliders, including finite width of the
  $W$'s and anomalous trilinear gauge boson couplings.
\item{} \textbf{Method of solution:} Monte Carlo event generation.
\item{} \textbf{Restrictions on the complexity of the problem:} The matrix
  elements for the hard subprocess contain only the doubly and singly
  resonant contributions constituting the so-called ``CC11'' diagrams,
  in the Born approximation.  The hadronization of final state quarks
  is handled by external programs.  Interfaces to the most popular QCD
  Monte Carlos are provided.
\item{} \textbf{Typical running time:} The time needed per unweighted
  event depends on the beam energy, cuts and CPU.  The test run (10000
  unweighted parton level events at LEP2 energies) takes approximately
  420 CPU seconds on an aging 486DX2/66 PC running Linux 2-$\epsilon$.
  On contemporary personal computers, the 42~ms/event will be reduced
  to 5~ms/event and work stations will need less than 1~ms/event.  The
  CPU time per event increases mildly with the beam energy.
\end{itemize}

%%%%%%%%%%%%%%%%%%%%%%%%%%%%%%%%%%%%%%%%%%%%%%%%%%%%%%%%%%%%%%%%%%%%%%%%
\newpage

\section{Introduction}
\label{sec:introduction}

The spectacular success of the high energy electron positron colliders
LEP at CERN and SLC at SLAC has confirmed the predictions of the
Standard Model (SM) for the interactions between the gauge bosons and
the fermions even at the level of electroweak radiative corrections.
The high precision of the experimental data allows to put limits on
parameters of yet unobserved particles like
the Higgs boson through the appearance of these particles in weak
loop corrections.  On the other hand, the non-Abelian structure of the
gauge sector of the SM predicts couplings between the electroweak gauge
bosons which have not been tested directly.  The forthcoming upgrade of
LEP~\cite{LEP86,AKV89} to LEP2~\cite{Aachen86,CERN-96-01}
with a center of mass energy of up to
$\sqrt{s} \approx 200$~GeV and a future 0.5--1.5~TeV $e^+e^-$ linear
collider~\cite{EE500} (NLC, for short) will in particular make the
trilinear $WW\gamma$ and $WWZ$
couplings directly observable through their contribution to the
production of $W$ pairs.

The main task of LEP2 will be a precise determination of the
mass and the width of the $W$ and the production cross section in the
threshold region~\cite{Delphi92}.  At NLC, a measurement of the total
cross
section, the $W$ angular distributions and in particular of their
longitudinal polarization component will give a handle on possible
anomalous couplings among the electroweak bosons induced by new physics
beyond the SM \cite{BKRS92,HPZH87}, and in turn reveal an insight into
the mechanism of electroweak symmetry breaking.

However, possible new physics is already severely constrained by present
LEP and SLC data, and the effects to be expected at NLC (even more
so at LEP2) are small.  In order to extract these small
effects, one has to have a precise knowledge of the radiative
corrections within the SM.

The electroweak radiative corrections to the production of on-shell
$W$'s to one-loop order are by now well established \cite{BDS+88,FJZ89}.
The influence of the finite width of the $W$'s has been investigated
in~\cite{MNW86,BBOR93}.  Also, the higher order QED corrections have
been calculated in the leading log approximation (LLA) and their
importance has been emphasized in ref.~\cite{CDMN91}.  An
exhaustive overview of the standard model predictions has been
provided in~\cite{BD94} and~\cite{CERN-96-01}.

Unfortunately, the experimental reconstruction of the $W$'s and the
determination of their polarization is complicated by the fact that they
may decay either into leptons with an escaping neutrino, or into
hadrons, where the jet energies may be poorly known due to undetected
particles.  In addition, the radiative corrections due to emission of
photons produce a systematic shift of the effective center of mass
energy towards smaller values.  Such effects may best be studied with
the help of a Monte Carlo event generator.

Although quite a few semianalytical calculations of the corrections
mentioned above have been available for some time, version 1.0 of
\WOPPER/ \cite{wopper-CPC} had been the first publicly available and
complete Monte Carlo event generator.  In the meantime, many new
programs have been released \cite{CERN-96-01}.

This paper describes version \Version/ of the Monte Carlo event
generator \WOPPER/.  It is based on the lowest order cross
section for the process $e^+ e^- \to (W^+W^-) \to 4$f and focusses on
QED radiative corrections in the LLA resummed to all orders in $\alpha$
and the effects of finite width of the $W$'s.  The four-momenta of the
exclusive hard photons are generated explicitly and treated with full
kinematics.  The $W$ decays into fermions are implemented at the parton
level, including leading QCD corrections to the $W$ branching fractions.

This write-up is organized as follows: In section~\ref{sec:general} we
outline the physics underlying the algorithms implemented in \WOPPER/.
The actual implementation is described in section~\ref{sec:prog-struct}.
The parameters controlling the execution of \WOPPER/ are discussed in
detail in section~\ref{sec:parameters} and the \f77/ interface is
presented in section~\ref{sec:f77}.
Distribution notes, a
listing of all external symbols and the output of a test run can be
found in the appendices.

%%%%%%%%%%%%%%%%%%%%%%%%%%%%%%%%%%%%%%%%%%%%%%%%%%%%%%%%%%%%%%%%%%%%

\section{Theoretical Background and General Features}
\label{sec:general}

%%%%%%%%%%%%%%%%%%%%%%%%%%%%%%%%%%%%%%%%%%%%%%%%%%%%%%%%%%%%%%%%%%%%

\subsection{QED corrections at very high energies}
\label{sec:formalism}

In the structure function formalism \cite{KF85,AM86,BBN89} the
expression for the radiatively corrected cross section reads
\begin{equation}
  \sigma(s) = \int\limits_0^1 dx_+ dx_- \;
  D(x_+,Q^2) D(x_-,Q^2) \; \hat\sigma(x_+ x_- s) \; ,
  \label{eq:factorization}
\end{equation}
where $\hat\sigma$ is the Born level cross section of the hard process,
$D(x,Q^2)$ are the structure functions for initial state radiation, and
$Q^2$ is the factorization scale.

The structure functions $D$ sum the numerically most important leading
logarithmic contributions
\begin{equation}
  \frac{\alpha}{\pi} \log \left(\frac{s}{m_e^2}\right)  \approx 6\%
  \qquad  \mbox{(at LEP2 and NLC energies)}
\end{equation}
to the electromagnetic radiative corrections to all orders.  They
satisfy the QED evolution equation~\cite{KF85}
\begin{eqnarray}
\label{eq:DGLAP}
   Q^2 \frac{\partial}{\partial Q^2} D(x,Q^2)
      & = & \frac{\alpha}{2\pi}
             \int\limits_x^1 \frac{dz}{z} \left[P_{ee}(z)\right]_+
                 D\left(\frac{x}{z},Q^2\right) \\
\label{eq:splitting-function}
  P_{ee}(z) & = & \frac{1+z^2}{1-z}
\end{eqnarray}
with initial condition
\begin{equation}
  D(x,m_e^2) = \delta(1-x) \; .
\end{equation}
The solution to eq.~(\ref{eq:DGLAP}) automatically includes the very
important exponentiation of the soft photon contributions to the
radiative corrections.

An explicitly regularized version of (\ref{eq:DGLAP})
used in the Monte Carlo implementation is given by
\begin{eqnarray}
\label{eq:regularized-DGLAP}
   Q^2 \frac{\partial}{\partial Q^2} D(x,Q^2)
     & = & \frac{\alpha}{2\pi}
              \int\limits_x^{1-\epsilon} \frac{dz}{z} P_{ee}(z)
                 D\left(\frac{x}{z},Q^2 \right) \\
     &   & \mbox{} - \frac{\alpha}{2\pi}
             \left[ \int\limits_0^{1-\epsilon} dz P_{ee}(z) \right]
                  D(x,Q^2). \nonumber
\end{eqnarray}
It is crucial to note that the characteristics of the generated event
sample do not depend on~$\epsilon$, if it is chosen well below the
experimental threshold for the detection of soft photons.  Theoretically,
the cross sections will remain positive for \emph{all} values of
$\epsilon$, but limited storage for ultra soft photons and limited
floating point range impose a lower limit.

The radiatively corrected cross section (\ref{eq:factorization}) is
implemented in a Monte Carlo event generator by solving the
integro-differential equation~(\ref{eq:regularized-DGLAP}) by iteration
and taking into account the energy loss in the hard cross section.  It
is very similar to algorithms for quark fragmentation in
QCD~\cite{Sjo85}.  As a by-product of the branching algorithm, the
four-momenta of the radiated photons are generated explicitly.

The initial state branching algorithm has already been used in the Monte
Carlo generator \texttt{KRONOS} \cite{ADM+92a} and an improved version in
the generator \texttt{UNIBAB} \cite{UNIBAB}, where the implementation has
been described in detail.

The momenta of the electron and positron after initial state branching
are then used as the input momenta for the subgenerator of the hard
subprocess described below.

%%%%%%%%%%%%%%%%%%%%%%%%%%%%%%%%%%%%%%%%%%%%%%%%%%%%%%%%%%%%%%%%%%%%

\subsection{Born Cross Section}
\label{sec:born}

In the general case, there are many Feynman diagrams contributing to the
process $ e^+e^- \to $ 4~fermions at high energies, even if one requires
that the quantum numbers of the final state fermions be consistent with
$W$ pair production \cite{CERN-96-01}.  However, the contribution of the
individual Feynman diagrams may be easily estimated by counting the
number of resonant propagators, where an intermediate vector boson may
come close to its mass shell.  Using this naive estimate, one finds that
the contribution of these `background diagrams' is suppressed by a
factor $\Gamma_W/M_W \sim $2.5\% for each non-resonant boson propagator,
and may be reduced further by appropriate cuts on the invariant masses
of the reconstructed $W$'s.  In fact, a full calculation~\cite{BKP94}
finds that this estimate is numerically justified for LEP2 energies,
unless electrons are in the final state and no invariant mass cuts are
applied.

In the current version \Version/ of \WOPPER/, we have taken into account
the doubly and singly resonant Feynman diagrams shown in
figures~\ref{fig:CC03-diagrams} and~\ref{fig:CC11-diagrams}, also
known as the ``CC11'' diagrams \cite{CERN-96-01}.  Note that the
combination of the fermions 1 and 2
has quantum numbers consistent with a $W^+$, while the combination of
the fermions 3 and 4 has quantum numbers consistent with a $W^-$, even
for the singly resonant diagrams.  We will refer to these pairs as
``pseudo-$W$'s'' below.

% Make this file work even without feynmf:
\expandafter\ifx\csname fmffile\endcsname\relax\else
\begin{fmffile}{manpics}
  % Compatibility hack for FeynMF versions _before_ 1.03, which are
  % still the defauft in some places:
  \expandafter\ifx\csname fmfpfx\endcsname\relax
    \let\nxp\noexpand  % version <= 1.02: halt expansion
  \else
    \let\nxp\relax     % version >= 1.03: hacks not needed anymore
  \fi
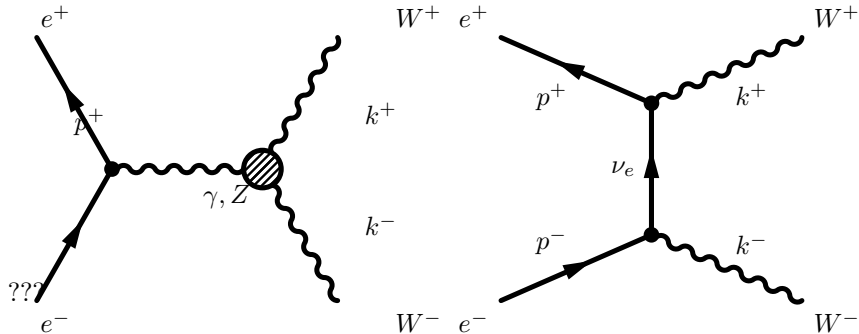
\begin{figure}
  \begin{center}
    \begin{fmfgraph*}(50,35)
      \fmfpen{thick}
      \fmfleftn{i}{2} \fmfrightn{o}{2}
      \fmf{fermion,lab=$p^-$,l.side=left}{i1,v1} \fmflabel{$e^-$}{i1}
      \fmf{fermion,lab=$p^+$,l.side=left}{v1,i2} \fmflabel{$e^+$}{i2}
      \fmf{photon,lab=$\gamma,,Z$}{v1,v2}
      \fmf{photon,lab=$k^-$,l.side=left}{v2,o1} \fmflabel{$W^-$}{o1}
      \fmf{photon,lab=$k^+$,l.side=right}{v2,o2} \fmflabel{$W^+$}{o2}
      \fmfblob{(.1w)}{v2}
      \fmfdot{v1}
    \end{fmfgraph*}
    \qquad\quad
    \begin{fmfgraph*}(50,35)
      \fmfpen{thick}
      \fmfleftn{i}{2} \fmfrightn{o}{2}
      \fmf{fermion,lab=$p^-$,l.side=left}{i1,v1} \fmflabel{$e^-$}{i1}
      \fmf{fermion,lab=$\nu_e$,l.side=left}{v1,v2}
      \fmf{fermion,lab=$p^+$,l.side=left}{v2,i2} \fmflabel{$e^+$}{i2}
      \fmf{photon,lab=$k^-$,l.side=left}{v1,o1} \fmflabel{$W^-$}{o1}
      \fmf{photon,lab=$k^+$,l.side=right}{v2,o2} \fmflabel{$W^+$}{o2}
      \fmfdotn{v}{2}
    \end{fmfgraph*}
  \end{center}
  \caption{%
    The Feynman diagrams contributing to the on-shell production
    $e^+e^- \to W^+ W^-$.} 
  \label{fig:on-shell-diagrams}
\end{figure}

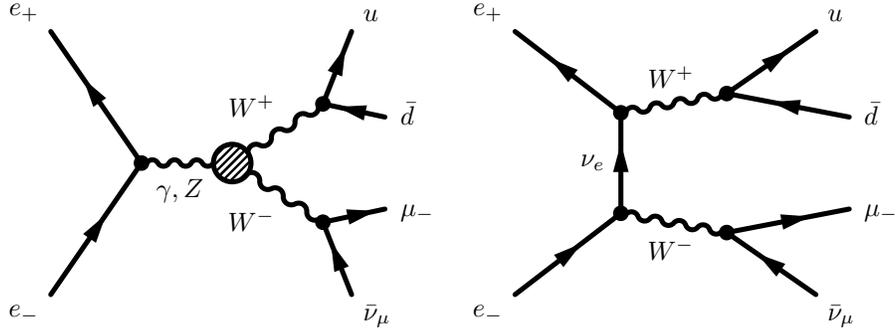
\begin{figure}
  \begin{center}
    \begin{fmfgraph*}(50,35)
      \fmfpen{thick}
      \fmfleftn{i}{2} \fmfrightn{o}{4}
      \fmflabel{$e_-$}{i1} \fmflabel{$e_+$}{i2}
      \fmflabel{$\nxp\bar\nu_{\mu}$}{o1}
      \fmflabel{$\mu_-$}{o2}
      \fmflabel{$\nxp\bar d$}{o3}
      \fmflabel{$u$}{o4}
      \fmf{fermion}{i1,v1,i2}
      \fmf{boson,lab=$\gamma,,Z$,tension=2}{v1,v2}
      \fmf{boson,lab=$W^-$,lab.side=right}{v2,v3}
      \fmf{fermion}{o1,v3,o2}
      \fmf{boson,lab=$W^+$,lab.side=left}{v2,v4}
      \fmf{fermion}{o3,v4,o4}
      \fmfdot{v1,v3,v4}\fmfblob{(.1w)}{v2}
    \end{fmfgraph*}
    \qquad\quad
    \begin{fmfgraph*}(50,35)
      \fmfpen{thick}
      \fmfleftn{i}{2} \fmfrightn{o}{4}
      \fmflabel{$e_-$}{i1} \fmflabel{$e_+$}{i2}
      \fmflabel{$\nxp\bar\nu_{\mu}$}{o1}
      \fmflabel{$\mu_-$}{o2}
      \fmflabel{$\nxp\bar d$}{o3}
      \fmflabel{$u$}{o4}
      \fmf{fermion}{i1,v1}
      \fmf{fermion}{v2,i2}
      \fmf{fermion,lab=$\nu_e$,lab.side=left}{v1,v2}
      \fmf{boson,lab=$W^-$,lab.side=right}{v1,v3}
      \fmf{fermion,tension=.5}{o1,v3,o2}
      \fmf{boson,lab=$W^+$,lab.side=left}{v2,v4}
      \fmf{fermion,tension=.5}{o3,v4,o4}
      \fmfdotn{v}{4}
    \end{fmfgraph*}
  \end{center}
  \caption{%
    The ``CC03'' subset of Feynman diagrams contributing to four
    fermion production $e^+e^- \to 4f$.}
  \label{fig:CC03-diagrams}
\end{figure}

\begin{figure}
  \begin{center}
    \begin{fmfgraph*}(50,35)
      \fmfpen{thick}
      \fmfleftn{i}{2} \fmfrightn{o}{4}
      \fmflabel{$e_-$}{i1} \fmflabel{$e_+$}{i2}
      \fmflabel{$\nxp\bar d$}{o1}
      \fmflabel{$u$}{o4}
      \fmflabel{$\nxp\bar\nu_{\mu}$}{o3}
      \fmflabel{$\mu_-$}{o2}
      \fmf{boson,lab=$\gamma,,Z$,tension=2}{v1,v2}
      \fmf{fermion}{i1,v1,i2} \fmf{fermion}{o1,v3,v2,o4}
      \fmffreeze
      \fmf{boson,lab=$W^-$}{v3,v4} \fmf{fermion}{o3,v4,o2}
      \fmfdotn{v}{4}
    \end{fmfgraph*}
    \qquad\quad
    \begin{fmfgraph*}(50,35)
      \fmfpen{thick}
      \fmfleftn{i}{2} \fmfrightn{o}{4}
      \fmflabel{$e_-$}{i1} \fmflabel{$e_+$}{i2}
      \fmflabel{$\nxp\bar\nu_{\mu}$}{o1}
      \fmflabel{$\mu_-$}{o4}
      \fmflabel{$\nxp\bar d$}{o3}
      \fmflabel{$u$}{o2}
      \fmf{boson,lab=$\gamma,,Z$,tension=2}{v1,v2}
      \fmf{fermion}{i1,v1,i2} \fmf{fermion}{o1,v3,v2,o4}
      \fmffreeze
      \fmf{boson,lab=$W^+$}{v3,v4} \fmf{fermion}{o3,v4,o2}
      \fmfdotn{v}{4}
    \end{fmfgraph*}\\
    \vspace*{2\baselineskip}
    \begin{fmfgraph*}(50,35)
      \fmfpen{thick}
      \fmfleftn{i}{2} \fmfrightn{o}{4}
      \fmflabel{$e_-$}{i1} \fmflabel{$e_+$}{i2}
      \fmflabel{$\nxp\bar d$}{o1}
      \fmflabel{$u$}{o4}
      \fmflabel{$\nxp\bar\nu_{\mu}$}{o3}
      \fmflabel{$\mu_-$}{o2}
      \fmf{boson,lab=$\gamma,,Z$,tension=2}{v1,v2}
      \fmf{fermion}{i1,v1,i2} \fmf{fermion}{o1,v2,v3,o4}
      \fmffreeze
      \fmf{boson,lab=$W^-$}{v3,v4} \fmf{fermion}{o3,v4,o2}
      \fmfdotn{v}{4}
    \end{fmfgraph*}
    \qquad\quad
    \begin{fmfgraph*}(50,35)
      \fmfpen{thick}
      \fmfleftn{i}{2} \fmfrightn{o}{4}
      \fmflabel{$e_-$}{i1} \fmflabel{$e_+$}{i2}
      \fmflabel{$\nxp\bar\nu_{\mu}$}{o1}
      \fmflabel{$\mu_-$}{o4}
      \fmflabel{$\nxp\bar d$}{o3}
      \fmflabel{$u$}{o2}
      \fmf{boson,lab=$Z$,tension=2}{v1,v2}
      \fmf{fermion}{i1,v1,i2} \fmf{fermion}{o1,v2,v3,o4}
      \fmffreeze
      \fmf{boson,lab=$W^+$}{v3,v4} \fmf{fermion}{o3,v4,o2}
      \fmfdotn{v}{4}
    \end{fmfgraph*}
  \end{center}
  \caption{%
    The ``CC11'' subset of Feynman diagrams contributing to four
    fermion production $e^+e^- \to q\bar q'q''\bar q'''$.
    For $q\bar q'\ell\nu_\ell$ and $\ell^-\bar\nu_\ell\ell^+\nu_\ell$
    final states only 10 and 9 diagrams contribute, respectively,
    because of the uncharged neutrinos in the intermediate states.}
  \label{fig:CC11-diagrams}
\end{figure}
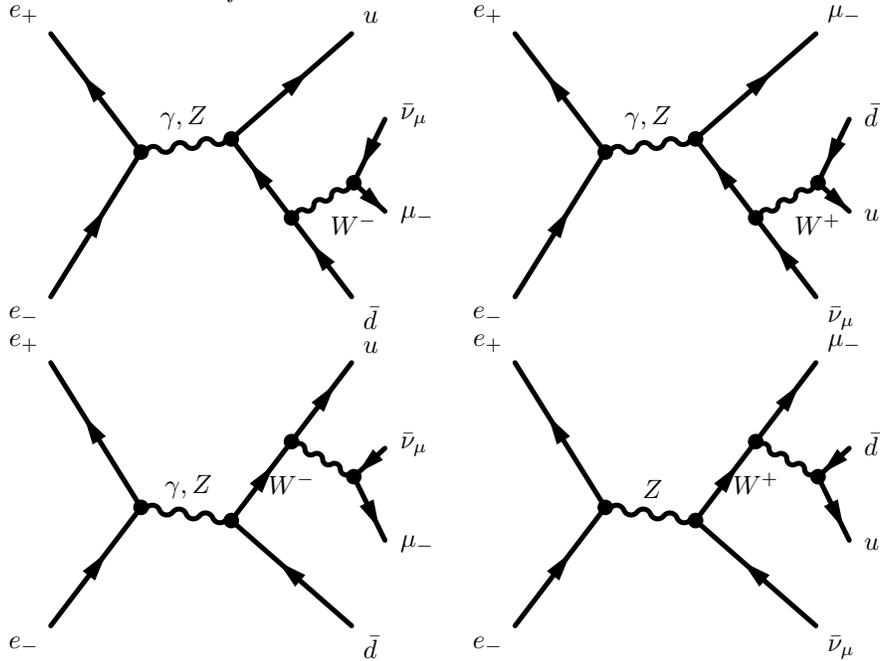
\end{fmffile}
\fi

In the calculation of the Born amplitudes, we use an energy dependent
width for both the $W$ and $Z$ propagators, that is given by the
contributions of the light fermions to the vector boson self energies:
\begin{equation}
  \mbox{Im } \Sigma_W(k^2) \equiv \sqrt{k^2} \cdot \Gamma_W(k^2)
  \approx \frac{k^2}{M_W^2} \; M_W \Gamma_W \; , \qquad
  \mbox{Re } \Sigma_W(k^2) \approx 0
  \label{eq:sigma}
\end{equation}
and similarly for $Z$'s.

It can easily be seen that, after integrating the modulus squared of the
Born amplitudes over the decay angles of the pseudo $W$'s, one obtains
the same resonance formula for the total cross section for production of
the pseudo $W$ pairs as in the well known doubly renonant case
\cite{MNW86,DS90,Aep91}, for each combination of final state fermions
$f_1\bar{f}_2 f_3\bar{f}_4$:
\begin{equation}
  \sigma (12,34) \; = \;
  \int ds_+ ds_- \;
  \frac{\sqrt{s_+} \,\Gamma_{W \to 12}(s_+)}{\pi D(s_+)}
  \frac{\sqrt{s_-} \,\Gamma_{W \to 34}(s_-)}{\pi D(s_-)}
  \sigma_{\mbox{off}}(s; s_+, s_-; 12,34)
  \label{eq:resonance-formula-partial}
\end{equation}
Here $\sigma_{\mbox{off}}(s; s_+, s_-)$ denotes the off-shell cross
section for the production of two pseudo-$W$'s, and $\Gamma_W(s_\pm)$ is
the effective decay width of a virtual $W$ into light fermions
according to eq.~(\ref{eq:sigma}).  The resonance factors are
Breit-Wigner functions with energy-dependent width:
\begin{equation}
  \frac{1}{D(s_\pm)} \; = \;
  \frac{1}{(s_\pm - M_W^2)^2 + s_\pm \Gamma_W^2(s_\pm)} \; .
\end{equation}
Note that equation (\ref{eq:resonance-formula-partial}) also holds for
the total cross section, after summation of over the decay channels of
the pseudo~$W$'s:
\begin{equation}
  \sigma \; = \;
  \int ds_+ ds_- \;
  \frac{\sqrt{s_+} \,\Gamma_{W}(s_+)}{\pi D(s_+)}
  \frac{\sqrt{s_-} \,\Gamma_{W}(s_-)}{\pi D(s_-)}
  \sigma_{\mbox{off}}(s; s_+, s_-)
  \label{eq:resonance-formula-total}
\end{equation}

For the actual implementation it is useful to apply the mappings
($\gamma \equiv \Gamma_W/M_W$)
\begin{equation}
  \xi_\pm =
  \arctan \left( \frac{(1+\gamma^2) s_\pm - M_W^2}{\gamma M_W^2} \right)
  \label{eq:mapping}
\end{equation}
to eq.~(\ref{eq:resonance-formula-total}) in order to get a smooth
integrand suitable for a Monte Carlo rejection algorithm:
\begin{eqnarray}
  \sigma  & = &
  \left. \int d\xi_+ d\xi_- \;\;
  \frac{1}{\pi} \frac{s_+}{M_W^2} \;
  \frac{1}{\pi} \frac{s_-}{M_W^2} \;
  \sigma_{\mbox{off}}(s; s_+, s_-)
  \right|_{s_\pm = M_W^2 \cdot (1 + \gamma\tan \xi_\pm) / (1+\gamma^2)}
  \nonumber \\  & \equiv &
  \int d\xi_+ d\xi_- \; \tilde\sigma(s; \xi_+, \xi_-)
  \label{eq:resonance-formula-mapped}
\end{eqnarray}

%%%%%%%%%%%%%%%%%%%%%%%%%%%%%%%%%%%%%%%%%%%%%%%%%%%%%%%%%%%%%%%%%%%%
\subsection{Triple gauge bosons couplings}
\label{sec:TGV}

In order to model the triple gauge boson vertices $WW\gamma$ and $WWZ$
including possible deviations from the Standard Model (``anomalous
couplings''), we use an effective Lagrangian as given by Hagiwara et
al.~\cite{HPZH87}:\footnote{%
  Here we use the conventions of Itzykson and Zuber with
  $\epsilon^{0123} = +1$.}
\begin{eqnarray}
\label{eq:L_WWV}
{\cal L}_{WWV}  /  g_{WWV}  & = &
i g_1^V \left( W^\dagger_{\mu\nu} W^\mu - W^\dagger_\mu W^\mu_{\;\;\;\nu}
\right) V^\nu +
i \kappa_V  W^\dagger_\mu W_\nu V^{\mu\nu}
\nonumber \\
&& + \frac{i \lambda_V}{m_W^2}
   W^\dagger_{\lambda\mu} W^\mu_{\;\;\;\nu} V^{\nu\lambda} -
g_4^V  W^\dagger_\mu W_\nu \left(\partial^\mu V^\nu + \partial^\nu V^\mu
\right)
\nonumber \\
&& + g_5^V \epsilon^{\mu\nu\lambda\sigma} \left( W^\dagger_\mu
\stackrel{\leftrightarrow}{\partial_\lambda} W_\nu \right) V_\sigma +
i \tilde\kappa_V  W^\dagger_\mu W_\nu \tilde{V}^{\mu\nu}
\nonumber \\
&& + \frac{i \tilde\lambda_V}{m_W^2}
   W^\dagger_{\lambda\mu} W^\mu_{\;\;\;\nu} \tilde{V}^{\nu\lambda}
\end{eqnarray}
Here $V^\mu$ stands for either the photon or the Z field, $W^\mu$ is the
$W^-$ field, $W_{\mu\nu} = \partial_\mu W_\nu - \partial_\nu W_\mu$,
$V_{\mu\nu} = \partial_\mu V_\nu - \partial_\nu V_\mu$, and
$\tilde{V}_{\mu\nu} = \frac{1}{2} \epsilon_{\mu\nu\lambda\sigma}
V^{\lambda\sigma}$.

These seven operators exhaust all possible Lorentz structures when we
neglect the scalar components of the vector bosons,
\begin{equation}
  \partial_\mu V^\mu = 0 \; , \qquad
  \partial_\mu W^\mu = 0 \; .
\end{equation}
which is allowed if we couple the bosons to (almost) massless external
fermions.

In accordance with Hagiwara et al.~\cite{HPZH87} we choose
\begin{equation}
  g_{WW\gamma} = -e \; , \qquad
  g_{WWZ}      = -e \cot \theta_W \; .
\end{equation}
In the Standard Model at tree level we then have:
\begin{equation}
  g_1^\gamma = g_1^Z = \kappa_\gamma = \kappa_Z = 1 \; ,
  \quad \mbox{all others couplings} = 0
\end{equation}

We demand P and C conservation separately for the $WW\gamma$ vertex,
thus we have to set
\begin{equation}
  g_4^\gamma = g_5^\gamma =
  \tilde\kappa_\gamma = \tilde\lambda_\gamma = 0 \; .
\end{equation}
Electromagnetic gauge invariance requires that $g_1^\gamma$ is related
to the charge of the W,
\begin{equation}
  g_1^\gamma = 1 \; ,
\end{equation}
so that we have only two free parameters ($\kappa_\gamma,
\lambda_\gamma$) for the $WW\gamma$ vertex, while there are seven for
the $WWZ$ vertex.

For convenience, the anomalous couplings can also be specified in
another popular parameterization
$(\delta_Z,x_\gamma,x_Z,y_\gamma,y_Z,z_Z)$. The translation formulae
are presented in section~\ref{sec:AC} below.

%%%%%%%%%%%%%%%%%%%%%%%%%%%%%%%%%%%%%%%%%%%%%%%%%%%%%%%%%%%%%%%%%%%%
\subsection{Coulomb singularity}
\label{sec:coulomb}

Another class of universal corrections, which is important near
threshold, is the so-called Coulomb singularity.  We implement the
correction~\cite{BBD93} 
\begin{equation}
  \sigma_{\scriptsize\rm Coulomb}
     = \sigma_{\scriptsize\rm Born} \frac{\alpha\pi}{2\beta}
      \left(1 - \frac{2}{\pi} \arctan\left(
        \frac{\vert\beta_M+\Delta|^2 - \beta^2}%
             {2\beta\mathop{\rm Im}\beta_M}\right)\right)
\end{equation}
to the off shell total cross section (\ref{eq:resonance-formula-total}).
Stricly speaking, this correction should only be applied to the subset
of doubly resonant diagrams.  However, the Coulomb correction is only
large for the $W$'s close to their mass shell where the singly resonant
diagrams are suppressed, and thus we apply it as a universal correction
factor.

%%%%%%%%%%%%%%%%%%%%%%%%%%%%%%%%%%%%%%%%%%%%%%%%%%%%%%%%%%%%%%%%%%%%%

\subsection{Decays of the pseudo $W$'s}
\label{sec:decay}

In the case of the doubly resonant diagrams, the relative probabilities
of the final state fermions are simply given by the branching fractions
of the $W$'s.  This remains true even after taking into account the
inclusive QCD corrections to the $W$ decays to quarks to first order in
$\alpha_S$, because of the factorization of the amplitude into $W$
production and decay.  Neglecting fermion masses, one has:
\begin{equation}
  \Gamma(W \to q \bar{q}') =
  N_C \left(1 + \frac{\alpha_S}{\pi} \right) \Gamma(W \to \ell \nu_\ell)
\end{equation}
However, in the case of the inclusion of the singly and doubly resonant
diagrams, it is not clear how to take the QCD corrections into account,
since no full calculation is as yet available.  Neglecting QCD
corrections to final states with quarks is clearly unacceptable, and
simply multiplying the contribution of the doubly resonant diagrams with
the QCD correction factor spoils the gauge cancellations between singly
and doubly resonant diagrams.  Therefore, we have taken the approach to
implement the QCD corrections in (\ref{eq:resonance-formula-partial}) by
using an ``effective number of colors''
$$
N_C^{\mbox{eff}} = N_C \left(1 + \frac{\alpha_S}{\pi} \right) \; ,
$$
which we refer to as the ``naive QCD corrections''.

The final state fermions of the decay $W \to f \bar f'$ are therefore
chosen with a probability according to the cross sections obtained from
(\ref{eq:resonance-formula-partial}), with the naive QCD corrections
applied as described above.  The branching fractions for hadronic decays
into~$u$, $d$, $c$, $s$, and~$b$~quarks are given by the corresponding
CKM matrix elements, while decays into~$t$~quarks are assumed to be
kinematically forbidden because of~$|m_t-M_W|\gg\Gamma_W$.  Currently,
the fermion masses are only taken into account kinematically and not in
the matrix elements.

The angular distributions of the pseudo $W$'s and of the decay fermions
is generated by a standard mapping and rejection algorithm from the
differential cross section $d\sigma(s;s_+,s_-;\theta;12,34)/d\Omega$ and
from the modulus squared matrix element, respectively.

%%%%%%%%%%%%%%%%%%%%%%%%%%%%%%%%%%%%%%%%%%%%%%%%%%%%%%%%%%%%%%%%%%%%%

\subsection{Hadronization}

If one or both of the $W$'s have decayed into quarks, they can
optionally be hadronized using either the LUND string
model~\cite{JETSET}, the HERWIG cluster model~\cite{MWA+92} or the
ARIADNE~\cite{ARIADNE} color dipole model.
In the present version it is not possible\footnote{%
  Unless one makes the necessary changes to the \WOPPER/ sources in
  routines \.{wwlund()}, \.{wwhwig()} and \.{wwaria()}.}
to study color-rearrangement effects in purely hadronic decays.  Both
$W$~decays are handled separately.

The hadronization model can be switched at run time with the parameter
\.{qcdmc}, which takes the values 0, 1, 2 and 3.  These correspond to
no parton level, \texttt{JETSET}~\cite{JETSET},
\texttt{HERWIG}~\cite{MWA+92} and ARIADNE~\cite{ARIADNE} respectively.

%%%%%%%%%%%%%%%%%%%%%%%%%%%%%%%%%%%%%%%%%%%%%%%%%%%%%%%%%%%%%%%%%%%%%

\section{Implementation of \WOPPER/ \Version/}
\label{sec:prog-struct}

Like almost all Monte Carlo event generators, \WOPPER/ is divided into
three parts: initialization, event generation, and termination.  These
are described in this section.  For ease of use, \WOPPER/ comes with two
application interfaces, so that a direct call to the lower level parts
in this section will never be necessary.  These interfaces will be
explained in section~\ref{sec:f77}.

%%%%%%%%%%%%%%%%%%%%%%%%%%%%%%%%%%%%%%%%%%%%%%%%%%%%%%%%%%%%%%%%%%%%

\subsection{Initialization}
\label{sec:initialization}

The initializations in \WOPPER/ are used for computing the value of
variables that will be used frequently during event generation.
Examples are the calculation of electroweak couplings, the maximum of
the off-shell Born cross section and other internal steering parameters
from the input parameters.  This is accomplished by a call to the
subroutine \.{wwinit} after setting the Monte Carlo parameters.  Since
\WOPPER/ does not yet include weak corrections, most of the
initializations performed are quite trivial.

Finally, a standard \hepevt/ initialization record \cite{AKV89} is
written, which may be used by the analysis program.

%%%%%%%%%%%%%%%%%%%%%%%%%%%%%%%%%%%%%%%%%%%%%%%%%%%%%%%%%%%%%%%%%%%%

\subsection{Event Generation}
\label{sec:generation}

The routine \.{wwgen} produces an event on every call.  The four momenta
of all generated particles as well as supplemental information is
written to a standard \hepevt/ event record, where it can be read by
user supplied analyzers.  See section~\ref{sec:hepevt} below for details
on \WOPPER/'s extensive use of \hepevt/.

The first step is the generation of the initial state radiation by the
branching routine \.{wwbini} using the algorithm of
section~\ref{sec:formalism} and of the virtual masses of the pseudo
$W$'s occurring in the intermediate state.  According to the cross
section (\ref{eq:resonance-formula-mapped}), the rejection weight is
calculated from the ratio of the actual off-shell cross section to the
maximum determined in the initialization step, and the event is accepted
with a probability according to this weight.

After the effective center of mass energy and the pseudo $W$ masses have
been fixed, the quantum numbers of the event (helicity of incoming
fermions and type of final state particles) are determined with
probabilities proportional to their relative cross sections.  The
angular distribution and four-momenta of the pseudo $W$'s are generated
in \.{wwgww} and \.{wwgppr}, respectively.  Finally, the decay of the
intermediate $W$'s into the final state fermions is accomplished in the
subroutine \.{wwgdec}.

%%%%%%%%%%%%%%%%%%%%%%%%%%%%%%%%%%%%%%%%%%%%%%%%%%%%%%%%%%%%%%%%%%%

\subsection{Termination}

The cross section for the generated events is obtained in the subroutine
\.{wwclos} from the standard formula
\begin{equation}
  \sigma_{tot}(s) =
  \max_{s';\xi_+,\xi_-} \{\tilde\sigma(s';\xi_+,\xi_-)\}
  \cdot \frac{\mbox{\# of successful trials}}{\mbox{total \# of trials}}
\end{equation}
where $\tilde\sigma$ is the off-shell cross section from
eq.~(\ref{eq:resonance-formula-mapped}).  $s'$ varies between 0 and $s$,
and $\xi_\pm$ in the range allowed by the specified cuts.  The
corresponding statistical error from the Monte Carlo integration is
given by
\begin{eqnarray}
  \Delta\sigma_{tot}(s)  & = &
  \max_{s';\xi_+,\xi_-} \{\tilde\sigma(s';\xi_+,\xi_-)\} \cdot \\
  && \sqrt{\frac{(\mbox{total \# of trials} - \mbox{\# of successful
   trials}) \cdot \mbox{\# of successful trials}}
   {(\mbox{total \# of trials})^3}} \nonumber
\end{eqnarray}
The cross section and the error on the cross section are placed into
\hepevt/, where they may be read by the user-supplied analyzer.

%%%%%%%%%%%%%%%%%%%%%%%%%%%%%%%%%%%%%%%%%%%%%%%%%%%%%%%%%%%%%%%%%%%%%

\subsection{Additional information in \hepevt/}
\label{sec:hepevt}

Because \WOPPER/ uses the standard \hepevt/ event record internally, not
only stable particles with \.{isthep(i) = 1} will be present.  Adapting
the conventions of the \HERWIG/ Monte Carlo \cite{MWA+92}, we use the
following status codes
\begin{itemize}
  \item{} 101: $e^-$ beam (positive $z$-direction),
  \item{} 102: $e^+$ beam (negative $z$-direction),
  \item{} 103: center of mass system of the collider,
  \item{} 110: $e^+e^-$ hard scattering center of mass system,
  \item{} 111: $e^-$ before hard scattering,
  \item{} 112: $e^+$ before hard scattering,
  \item{} 113: virtual $W^-$ after hard scattering,
  \item{} 114: virtual $W^+$ after hard scattering.
\end{itemize}
However, these entries have \emph{no} physical significance and should
\emph{never} be used in any analysis (an exception to this rule are the
beam particles 101 and 102, which are convenient for defining the
reference frame and are used e.g.~by the analyzer \hepawk/~\cite{Ohl92a}
for this purpose).  Only the particles with status code 1 belong to the
final state as predicted by \WOPPER/.

If no hadronization Monte Carlo is active, final state quarks will be
entered as ``stable'' particles with status code 1.

%%%%%%%%%%%%%%%%%%%%%%%%%%%%%%%%%%%%%%%%%%%%%%%%%%%%%%%%%%%%%%%%%%%%%%%%

\section{Parameters}
\label{sec:parameters}

The parameters controlling \WOPPER/ version \Version/ are summarized
in tables~\ref{tab:wopper-physics-parm}, \ref{tab:wopper-ac-parm}
and~\ref{tab:wopper-technical-parm}.  They will be described in the 
following subsections. %%% Their names should be almost self-explaining.

\begin{table}
  \begin{minipage}{\textwidth}
  \begin{center}
  \begin{tabular}{@{\tablestrut}|c|c|c|}
    \hline\hline
    Variable name   & semantics             & Default value
    \\\hline\hline
    \.{ebeam}       & $e^+$/$e^-$ beam energy & 250 GeV
    \\\hline
    \.{epol}        & longitudinal $e^-$ beam polarization & 0
    \\\hline
    \.{ppol}        & longitudinal $e^+$ beam polarization & 0
    \\\hline
    \.{scheme}      & renormalization scheme  & 1
    \\\hline
    \.{ahpla}       & $1/\alpha_{QED}(4M_W^2)$ & 128
    \\\hline
    \.{ahpla0}      & $1/\alpha_{QED}(0)$   & 137.0359895
    \\\hline
    \.{mass1e}      & $m_{e^\pm}$           &
                                         $0.51099906\cdot 10^{-3}$ GeV
    \\\hline
    \.{mass1w}      & $M_{W^\pm}$           & 80.26  GeV
    \\\hline
    \.{gamm1w}      & $\Gamma_{W}$          &  2.08  GeV
    \\\hline
    \.{mass1z}      & $M_{Z^0}$             & 91.1884 GeV
    \\\hline
    \.{gamm1z}      & $\Gamma_{Z}$          &  2.492 GeV
    \\\hline
    \.{gfermi}      & $G_F$                 &  $1.16639\cdot 10^{-5}$ GeV
    \\\hline
    \.{sin2w}       & $\sin^2\theta_W$      &  0.2310
    \\\hline
    \.{alphas}      & $\alpha_{QCD}(M_W^2)$ &  0.12
    \\\hline
    \.{cc11}        & all CC11 diagrams     &  \texttt{.true.}
    \\\hline
    \.{floops}      & fermion loop contributions &  \texttt{.false.}
    \\\hline
    \.{coulom}      & Include final state coulomb corrections & \texttt{.false.}
    \\\hline
    \.{ckmvus}      & $V_{us}=\sin\theta_C$ &  0.2196
    \\\hline
    \.{ckmvcb}      & $V_{cb}$              &  0.0400
    \\\hline
    \.{ckmvub}      & $V_{ub}$              &  0.0032
    \\\hline
  \end{tabular}
  \end{center}
  \end{minipage}
  \caption{Standard model parameters controlling \WOPPER/.}
  \label{tab:wopper-physics-parm}
\end{table}

\begin{table}
  \begin{minipage}{\textwidth}
  \begin{center}
  \begin{tabular}{@{\tablestrut}|c|c|c|}
    \hline\hline
    Variable name   & semantics             & Default value
    \\\hline\hline
    \.{acpara}      & parameterization      & 0
    \\\hline
    \.{kappag}      & $\kappa_\gamma$       & 1
    \\\hline
    \.{lamdag}      & $\lambda_\gamma$      & 0
    \\\hline
    \.{kappaz}      & $\kappa_Z$            & 1
    \\\hline
    \.{lamdaz}      & $\lambda_Z$           & 0
    \\\hline
    \.{g1z}         & $g_1^Z$               & 1
    \\\hline
    \.{g4z}         & $g_4^Z$               & 0
    \\\hline
    \.{g5z}         & $g_5^Z$               & 0
    \\\hline
    \.{kapzt}       & $\tilde\kappa_Z$      & 0
    \\\hline
    \.{lamzt}       & $\tilde\lambda_Z$     & 0
    \\\hline
    \.{acdz}        & $\delta_Z$            & 0
    \\\hline
    \.{acxg}        & $x_\gamma$            & 0
    \\\hline
    \.{acxz}        & $x_Z$                 & 0
    \\\hline
    \.{acyg}        & $y_\gamma$            & 0
    \\\hline
    \.{acyz}        & $y_Z$                 & 0
    \\\hline
    \.{aczz}        & $z_Z$                 & 0
    \\\hline
  \end{tabular}
  \end{center}
  \end{minipage}
  \caption{Anomalous couplings parameters controlling \WOPPER/.}
  \label{tab:wopper-ac-parm}
\end{table}

\begin{table}
  \begin{minipage}{\textwidth}
  \begin{center}
  \begin{tabular}{@{\tablestrut}|c|c|c|}
    \hline\hline
    Variable name   & semantics             & Default value
    \\\hline\hline
    \.{cutmin}      & minimum $W^\pm$ virtuality & 0 GeV
    \\\hline
    \.{cutmax}      & maximum $W^\pm$ virtuality & $\sqrt{s} = 2 E_{Beam}$
    \\\hline
    \.{nevent}      & Number of events      & 10000
    \\\hline
    \.{cc}          & apply canonical cuts  & .false.
    \\\hline
    \.{cclvl}       & level of canonical cuts  & 0
    \\\hline
    \.{ccver}       & version of canonical cuts  & 1
    \\\hline
    \.{bstyle}      & Key for QED radiation & 1
    \\\hline
    \.{circe}       & beamstrahlung         &  \.{.false.}
    \\\hline
    \.{ciracc}      & accelerator           &  2 (TESLA)
    \\\hline
    \.{cirver}      & version               &  1
    \\\hline
    \.{cirrev}      & revision              &  19960401
    \\\hline
    \.{circht}      & chattines             &  1
    \\\hline
    \.{epsiln}      & Internal infrared cutoff & $10^{-5}$
    \\\hline
    \.{qcdmc}       & Key for QCD Monte Carlo  & 0
    \\\hline
    \.{rangen}      & Random number generator  & 1
    \\\hline
    \.{rseed}       & Random number seed       & 54217137
    \\\hline
    \.{rlux}        & `Luxury' of generator    & 3
    \\\hline
    \.{errmax}      & maximum error count      & 100
    \\\hline
    \.{verbos}      & verbosity                & 0
    \\\hline
    \.{runid}       & run identification       &
    \\\hline
    \.{stdin}       & standard input           & 5
    \\\hline
    \.{stdout}      & standard output          & 6
    \\\hline
    \.{stderr}      & standard error           & 6
    \\\hline
  \end{tabular}
  \end{center}
  \end{minipage}
  \caption{Technical parameters controlling \WOPPER/.}
  \label{tab:wopper-technical-parm}
\end{table}

%%%%%%%%%%%%%%%%%%%%%%%%%%%%%%%%%%%%%%%%%%%%%%%%%%%%%%%%%%%%%%%%%%%%
\subsection{Electroweak Parameters}

Since the present version of \WOPPER/ does not include an electroweak
library, the electroweak parameters, namely the masses of the
electroweak bosons (\.{mass1w}, \.{mass1z}), their widths (\.{gamm1w},
\.{gamm1z}) and the Weinberg angle (\.{sin2w}) are treated as
independent parameters.  They enter an effective Born cross section and
may be set directly by the user.  As a special case, setting \.{gamm1w}
to 0 reproduces the narrow-width approximation with on-shell $W$'s in
the intermediate state.

The bulk of the non-electromagnetic radiative corrections can be
incorporated into the hard cross section by using the running QED
coupling $\alpha_{QED}(4M_W^2) \approx 1/128$ at the $W$ scale.  This value
is, however, \emph{not} correct for the initial state radiation of
on-shell photons, where $\alpha_{QED}(0) \approx 1/137$ has to be
used.  The inverse of the former value can be changed with \.{ahpla}
and that of the latter with \.{ahpla0}.

The presence of initial state radiation can be toggled using the
parameter \.{bstyle}.  The supported supported values are 0 and 1,
corresponding to no QED radiative corrections and LLA resummed initial
state QED radiative corrections.

The variable \.{cc11} controls the number of Feynman diagrams taken into
account.  The default value is \.{.true.}, which amounts to using all
diagrams shown in figures~\ref{fig:CC03-diagrams}
and~\ref{fig:CC11-diagrams}.  Setting this variable to
false restricts the calculation to only the doubly resonant (``signal'')
diagrams in figure~\ref{fig:CC03-diagrams}, also known as ``CC03''
diagrams.

As a matter of convention, all gauge boson widths are treated in
\WOPPER/ as running widths (c.f.\ eq.~(\ref{eq:sigma})).  However, it is
well known that the inclusion of an energy-dependence in the width of
the gauge bosons distorts the cancellations between the photon and
$Z$-exchange contributions of the signal diagrams.  It has been noted
\cite{Arg95} that the cancellations may be restored by including a
minimal set of higher-order diagrams, namely the fermion loop
contributions to the $ZWW$ vertex.  Neglecting the masses of all
fermions except for the top-quark which is assumed to be very heavy,
this so-called ``fermion loop scheme'' amounts to replacing the Standard
Model $ZWW$ tree-level vertex by:
\begin{equation}
 \Gamma^{ZWW} \to \Gamma^{ZWW}_{\mbox{\scriptsize floops}} =
 \Gamma^{ZWW}_{\mbox{\scriptsize tree}}
 \cdot \left(1 + i \frac{\Gamma_Z}{M_Z} \right)
\end{equation}
The inclusion of this factor is controlled by the variable \.{floops}.

Since version 1.3 it is possible to choose canonical input parameters
for benchmarking LEP2 Monte Carlos~\cite{CERN-96-01} by changing the
value of \.{scheme}. The following values are supported:
\begin{itemize}
  \item{} \.{scheme = 0}: ``free scheme'', all parameters are taken
    from the input and treated as independent parameters.
  \item{} \.{scheme = 1} (default): ``$G_F$ scheme''
    \begin{eqnarray}
      \sin^2\theta_W & = &
        \frac{\pi\alpha_{QED}(4M_W^2)}{\sqrt{2} G_F M_W^2} \\
      \Gamma_W & = & \frac{G_F M_W^3}{\sqrt{8}\pi}
                     \left(3+ \frac{2\alpha_{QCD}}{\pi}\right)
    \end{eqnarray}
  \item{} \.{scheme = 2}: ``$\sin\theta_W^{\textrm{eff.}}$ scheme''
    \begin{eqnarray}
      G_F & = &
        \frac{\pi\alpha_{QED}(4M_W^2)}{\sqrt{2} \sin^2\theta_W M_W^2} \\
      \Gamma_W & = & \frac{G_F M_W^3}{\sqrt{8}\pi}
                     \left(3+ \frac{2\alpha_{QCD}}{\pi}\right)
    \end{eqnarray}
  \item{} \.{scheme = 3}: ``Born scheme'', tree level formulas,
    independent $\sin\theta_W$, $G_F$,
    $\alpha_{QED}(4M_W^2)=\alpha_{QED}(0)$ and
    \begin{eqnarray}
      \Gamma_W & = & \frac{G_F M_W^3}{\sqrt{8}\pi}
                     \left(3+ \frac{2\alpha_{QCD}}{\pi}\right)
    \end{eqnarray}
  \item{} \.{scheme = -1, -2, -3}: same as the positive values,
    except for~$\Gamma_W$ which is not calculated from the standard
    model expression but taken from \.{gamm1w} instead.
\end{itemize}
        
%%%%%%%%%%%%%%%%%%%%%%%%%%%%%%%%%%%%%%%%%%%%%%%%%%%%%%%%%%%%%%%%%%%%
\subsection{Anomalous couplings}
\label{sec:AC}

The parameterization of anomalous $WWZ$ and $WWZ$ couplings in
table~\ref{tab:wopper-ac-parm} is controlled by the \.{acpara}
variable:
\begin{itemize}
  \item{} \.{acpara = 0}: no anomalous couplings at all.  During
    initialization, $\kappa_\gamma$, $\lambda_\gamma$, $\kappa_Z$,
    $\lambda_Z$, $g_1^Z$, $g_4^Z$, $g_5^Z$, $\tilde\kappa_Z$,
    $\tilde\lambda_Z$ will be reset to their standard model values.
  \item{} \.{acpara = 1}: anomalous couplings in the Hagiwara et
    al.~\cite{HPZH87} parameterization, using the values of
    $\kappa_\gamma$, $\lambda_\gamma$, $\kappa_Z$, $\lambda_Z$,
    $g_1^Z$, $g_4^Z$, $g_5^Z$, $\tilde\kappa_Z$, $\tilde\lambda_Z$
    specified in the variables from table~\ref{tab:wopper-ac-parm}.
  \item{} \.{acpara = 2}: anomalous couplings in the
    $(\delta_Z,x_\gamma,x_Z,y_\gamma,y_Z,z_Z)$ parameterization.
    The corresponding Hagiwara et al.~parameters
    \begin{eqnarray}
      \kappa_\gamma & = & 1 + x_\gamma \\
      \lambda_\gamma & = & y_\gamma \\
      \kappa_Z & = & 1 + \left(x_Z+\delta_Z\right) \tan\theta_W \\
      \lambda_Z & = & y_Z \tan\theta_W \\
      g_1^Z & = & 1 + \delta_Z \tan\theta_W \\
      g_4^Z & = & 0 \\
      \label{eq:g5Z}
      g_5^Z & = & - \frac{s}{M_W^2} z_Z \tan\theta_W \\
      \tilde\kappa_Z & = & 0 \\
      \tilde\lambda_Z & = & 0
    \end{eqnarray}
    will be used in the Monte Carlo.
\end{itemize}
The relation~(\ref{eq:g5Z}) between $g_5^Z$ and~$z_Z$ is only
approximate.  $z_Z$ is defined as the coefficient in the higher
dimensional operator
\begin{equation}
  \mathcal{L}_{\not C \not P}
     = \frac{ez_Z}{M_W^2} \partial_\mu \tilde Z_{\nu\rho}
         \left( \partial^\nu W^{-\rho} W^{+\mu}
                - \partial^\nu W^{-\mu} W^{+\rho}
                + \partial^\nu W^{+\rho} W^{-\mu}
                - \partial^\nu W^{+\mu} W^{-\rho} \right)
\end{equation}
In the approximation of massless final state fermions and equal
invariant mass of the intermediate~$W$'s, the relation~(\ref{eq:g5Z})
can be shown.  Radiative corrections cause a further deviation,
because of the change in~$s$.  Until future version of \WOPPER/ will
support~$z_Z$ directly, (\ref{eq:g5Z}) is an excellent approximation.

It should be noted that even a small value of~$z_Z$ will have a strong
effect at NLC energies, because of the $s$-factor in (\ref{eq:g5Z})
induced by the additional mass dimensions
in~$\mathcal{L}_{\not C \not P}$.  The natural value of~$z_Z$ will
therefore be much smaller than that of the other parameters.

%%%%%%%%%%%%%%%%%%%%%%%%%%%%%%%%%%%%%%%%%%%%%%%%%%%%%%%%%%%%%%%%%%%%
\subsection{Cuts}

In the present version \Version/ of \WOPPER/, only cuts in the
virtualities of the intermediate $W$'s are implemented in the event
generation:
\begin{itemize}
  \item{} \.{cutmin}: minimum virtuality of the intermediate $W^\pm$s,
  \item{} \.{cutmax}: maximum virtuality of the intermediate $W^\pm$s.
\end{itemize}
The cuts in virtualities have to satisfy the following conditions:
\begin{equation}
  0 \le \mbox{\.{cutmin}} < \mbox{\.{cutmax}} \le \sqrt{s} = 2 E_{Beam}
  \label{eq:cut-conditions}
\end{equation}
A value of 0 for \.{cutmax} will automatically be reset to the available
maximum, namely $2 E_{Beam}$.

Since version 1.3 it is possible to apply the canonical cuts for
benchmarking LEP2 Monte Carlos~\cite{CERN-96-01} to the event record by
setting \.{cc} to \.{.true.}.  These cuts will be reflected in the
caclulated total cross section.

%%%%%%%%%%%%%%%%%%%%%%%%%%%%%%%%%%%%%%%%%%%%%%%%%%%%%%%%%%%%%%%%%%%%
\subsection{Beamstrahlung}

\.{WOPPER} can be linked with the \.{CIRCE} library~\cite{CIRCE} to
include the effects of the beam-beam interaction at a linear
collider.  If the parameter \.{circe} is set to \.{.true.}, the
momenta of the incoming electrons and positrons will be distributed
according to the version~(\.{cirver}) and revision~(\.{cirver}) of the
\.{CIRCE} parameterization of a particular accelerator design
class(\.{ciracc}). 

%%%%%%%%%%%%%%%%%%%%%%%%%%%%%%%%%%%%%%%%%%%%%%%%%%%%%%%%%%%%%%%%%%%%
\subsection{Monte Carlo Parameters}

The remaining, more technical Monte Carlo parameters should be almost
self explaining.  Since our branching algorithm automatically includes
soft photon exponentiation, the results will not depend on the value of
the internal infrared cutoff \.{epsiln} (which is measured in units of
the beam energy), provided it is kept \emph{well below} the
experimental energy resolution.  However, it is not advisable to set it
many orders of magnitude lower than the default value, because this may
result in too high photon multiplicities that will overflow internal
tables.

A note on the random number generators available: the default value~1 of
the parameter \.{rangen} corresponds to the standard
\.{RANMAR}~\cite{MZT90} generator, that has been the generator of choice
for quite some time.  Recently, the quality of the random numbers
generated by \.{RANMAR} has been questioned and unwanted correlations
have been found, that caused large systematic errors in solid state
physics simulations~\cite{FLW92}.  A superior variation \.{RANLUX} has
been proposed~\cite{Lue93}, which is however \emph{much} slower.
If \WOPPER/ has been linked with the \.{CERN} library, 
setting \.{rangen} to~2 switches to \.{RANLUX} which can be used at the
5 ``luxury levels'' 0 to~4.  At the highest ``luxury levels'',
\.{RANLUX} will be rather slow, however.

However, since event generation involves a lot of decisions that
effectively randomize the subsequences used by themselves, we do not
expect that the correlations in \.{RANMAR} have any significant effect
on the event samples generated by \WOPPER/.  The \.{RANLUX} option has
been added to \WOPPER/ for some experimentation only.  It is left in
only because there is no particular reason for throwing it out again.

An even more interesting option is the random number generator
recently proposed by Donald Knuth~\cite{Knuth:TAO:errata}.  This
generator is selected by setting \.{rangen} to~3.  It generates 30-bit
integers with the following desirable properties:
\begin{itemize}
  \item the bit stream passed all the tests from George Marsaglia's
    ``diehard'' suite of tests for random number
    generators~\cite{Mar96} (the result of the ``birthday-spacing''
    test has to used with great care, because the position of the
    least significant bit of the 30-bit words varies if the bit stream
    in distributed into 32-bit words).
  \item it can be implemented with portable signed 32-bit arithmetic
    (\.{Fortran} can't do unsigned arithmetic).
  \item it can create at least $2^{30}-2$ independent sequences, which
    are selected by a 30-bit integer seed value.
  \item it is very fast.
\end{itemize}
The default value of \.{rangen} is still~1, but all users are
encouraged to give the Knuth generator a try.

%%%%%%%%%%%%%%%%%%%%%%%%%%%%%%%%%%%%%%%%%%%%%%%%%%%%%%%%%%%%%%%%%%%%

\subsection{QCD Parameters}

The parameters for \.{JETSET} should be accessed trough \WOPPER/'s
\.{lugive} command, which is translated directly to \.{JETSET}'s
\.{LUGIVE()} subroutine.  See the \.{JETSET} manual~\cite{JETSET} for
a comprehensive description of the available parameters and their
effects.

Since \.{HERWIG} does not sport the equivalent of the \.{LUGIVE()}
routine, its parameters have to be accessed through the standard
\WOPPER/ access mechanisms.  Tables~\ref{tab:wopper-herwig-parm}
and~\ref{tab:wopper-technical-herwig-parm} provide a list of the
available \.{HERWIG} parameters and the names under whith they are known
to \WOPPER/.  See the \.{HERWIG} manual~\cite{MWA+92} for a
comprehensive description of the effects of these parameters.

\begin{table}
  \begin{minipage}{\textwidth}
  \begin{center}
  \begin{tabular}{@{\tablestrut}|c|c|c|c|}
    \hline\hline
    \WOPPER/   & \.{HERWIG}   & semantics   & Default value
    \\\hline\hline
    \.{hwqcdl} & \.{QCDLAM}   & $\Lambda_{QCD}/\mathop{\rm GeV}$
                                            & 0.18
    \\\hline
    \.{hwrms1} & \.{RMASS(1)} & $m_d/\mathop{\rm GeV}$
                                            & 0.32
    \\\hline
    \.{hwrms2} & \.{RMASS(2)} & $m_u/\mathop{\rm GeV}$
                                            & 0.32
    \\\hline
    \.{hwrms3} & \.{RMASS(3)} & $m_s/\mathop{\rm GeV}$
                                            & 0.5
    \\\hline
    \.{hwrms4} & \.{RMASS(4)} & $m_c/\mathop{\rm GeV}$
                                            & 1.8
    \\\hline
    \.{hwrms5} & \.{RMASS(5)} & $m_b/\mathop{\rm GeV}$
                                            & 5.2
    \\\hline
    \.{hwrms6} & \.{RMASS(6)} & $m_t/\mathop{\rm GeV}$
                                            & 100.00
    \\\hline
    \.{hwrms0} & \.{RMASS(13)} & $m_g^{eff.}/\mathop{\rm GeV}$
                                            & 0.75
    \\\hline
    \.{hwvqcu} & \.{VQCUT}    & Quark virtuality cutoff
                                            & 0.48
    \\\hline
    \.{hwvgcu} & \.{VGCUT}    & Gluon virtuality cutoff
                                            & 0.10
    \\\hline
    \.{hwvpcu} & \.{VPCUT}    & Photon virtuality cutoff
                                            & -1.00
    \\\hline
    \.{hwclma} & \.{CLMAX}    & Max.~cluster mass parameter
                                            & 3.35
    \\\hline
    \.{hwpspl} & \.{PSPLT}    & Split cluster parameter
                                            & 1.00
    \\\hline
    \.{hwqdiq} & \.{QDIQK}    & Max.~scale for $g\to\mathop{\rm diquark}$
                                            & 0.00
    \\\hline
    \.{hwpdiq} & \.{PDIQK}    & $g\to\mathop{\rm diquark}$ rate parameter
                                            & 5.00
    \\\hline
    \.{hwqspa} & \.{QSPAC}    & Spacelike evolution cutoff
                                            & 2.50
    \\\hline
    \.{hwptrm} & \.{PTRMS}    & Intrinsic $p_T$
                                            & 0.00
    \\\hline
  \end{tabular}
  \end{center}
  \end{minipage}
  \caption{\.{HERWIG} parameters accessible from \WOPPER/.}
  \label{tab:wopper-herwig-parm}
\end{table}

\begin{table}
  \begin{minipage}{\textwidth}
  \begin{center}
  \begin{tabular}{@{\tablestrut}|c|c|c|c|}
    \hline\hline
    \WOPPER/   & \.{HERWIG}   & semantics   & Default value
    \\\hline\hline
    \.{hwipri} & \.{IPRINT}   & Printout option
                                            & 1
    \\\hline
    \.{hwmaxp} & \.{MAXPR}    & Max. printouts
                                            & 0
    \\\hline
    \.{hwmaxe} & \.{MAXER}    & Max. errors
                                            & 10
    \\\hline
    \.{hwlwev} & \.{LWEVT}    & Event output unit
                                            & 0
    \\\hline
    \.{hwlrsu} & \.{LRSUD}    & Sudakov input unit
                                            & 0
    \\\hline
    \.{hwlwsu} & \.{LWSUD}    & Sudakov output unit
                                            & 77
    \\\hline
    \.{hwsudo} & \.{SUDORD}   & Sudakov order in $\alpha_S$
                                            & 1
    \\\hline
    \.{hwnrn1} & \.{NRN(1)}   & First random seed
                                            & 17673
    \\\hline
    \.{hwnrn2} & \.{NRN(2)}   & Second random seed
                                            & 63565
    \\\hline
    \.{hwazso} & \.{AZSOFT}   & Soft gluon azimuthal correlations
                                            & .true.
    \\\hline
    \.{hwazsp} & \.{AZSPIN}   & Gluon spin azimuthal correlations
                                            & .true.
    \\\hline
    \.{hwb1li} & \.{B1LIM}   & $B$-cluster $\to$ 1 hadron parameter
                                            & 0.0
    \\\hline
  \end{tabular}
  \end{center}
  \end{minipage}
  \caption{Technical \.{HERWIG} parameters accessible from \WOPPER/.}
  \label{tab:wopper-technical-herwig-parm}
\end{table}

\.{ARIADNE} doesn't supply a high level parameter changing interface
(like \.{LUGIVE()}) as well.  Since the tunable parameters are more
copious than in the case of \.{HERWIG}, we have refrained from making
them accessible through the \WOPPER/ parameter interface.  Users who
wish to change \.{ARIADNE} paramters are invited to add a call to
their parameter changing routine to \.{wwaria()}.  The standard
\.{ARIADNE} tuning sets can however be accessed with the \.{artune}
command, which passes its argument to the \.{ARTUNE()} routine.

%%%%%%%%%%%%%%%%%%%%%%%%%%%%%%%%%%%%%%%%%%%%%%%%%%%%%%%%%%%%%%%%%%%%
\section{\f77/ Interface}
\label{sec:f77}

\WOPPER/ version \Version/ provides two application program interfaces
on different levels.  The higher (preferred) level consists of the
command interpreter \.{wwdcmd} that accepts commands in form of
\.{character*(*)} strings.  This driver communicates with the analyzer
\hepawk/ \cite{Ohl92a} by default.  The lower level consists of two \f77/
subroutine calls: \.{wwpsrv} and \.{wopper}.

%%%%%%%%%%%%%%%%%%%%%%%%%%%%%%%%%%%%%%%%%%%%%%%%%%%%%%%%%%%%%%%%%%%%
\subsection{Higher Level Interface}
\label{sec:f77-high-level}

\begin{example}{Higher level \f77/ interface}{ex:f77}
\.{* wopperappl.f}                                                   \\
\>\>\> \.{...}                                                       \\
\>\>\> \.{call wwdcmd ('set ebeam 95')}   \C select $\sqrt s=190$GeV \\
\>\>\> \.{call wwdcmd ('set acpara 2')} \C select anomalous coupling \\
\>\>\> \.{call wwdcmd ('set acdz 0.01')}          \C $\delta_Z=0.01$ \\
\>\>\> \.{...}                                                       \\
\>\>\> \.{call wwdcmd ('init')}          \C initialize the generator \\
\>\>\> \.{call wwdcmd ('generate 10000')}   \C generate 10000 events \\
\>\>\> \.{...}                                                       \\
\>\>\> \.{call wwdcmd ('close')}                          \C cleanup \\
\>\>\> \.{...}
\end{example}

The simple commands understood by \.{wwdcmd} are  (here keywords are
typeset in typewriter font and variables in italics; vertical bars
denote alternatives)
\begin{itemize}
  \item \.{initialize} \hfil\goodbreak
    Force initialization of \WOPPER/ and write an initialization
    record into the
    \hepevt/ event record, which should trigger the necessary
    initializations in the analyzer.
  \item \.{generate} $[n]$ \hfil\goodbreak
    Generate \.{nevent} events and call \hepawk/ to analyze them.
    If the optional parameter $n$ is supplied, \.{nevent} is set
    to its value.
  \item \.{close} \hfil\goodbreak
    Write a termination record to \hepevt/, which should
    trigger the necessary cleanups in the analyzer.
  \item \.{statistics} \hfill\goodbreak
    Print performance statistics (this is usually only useful for the
    \WOPPER/ developers, who are tuning internal parameters).
  \item \.{quit} \hfil\goodbreak
    Terminate \WOPPER/ without writing a termination
    record.
  \item \.{exit}$\vert$\.{bye} \hfil\goodbreak
    Write a termination record and terminate \WOPPER/.
  \item \.{set} \textit{variable}
         \textit{ival}$\vert$\textit{rval} \hfil\goodbreak
    Set physical or internal parameters.  See the tables
    \ref{tab:wopper-physics-parm} and \ref{tab:wopper-technical-parm}
    for a comprehensive listing of all variables.  For example, the
    command \verb+'set ahpla 127.0'+ will set the QED fine structure
    constant to $1/127$.
  \item \.{print} \textit{variable}$\vert$\.{all} \hfil\goodbreak
    Print the value of physical or internal variables.
    \.{print *} causes a listing of all variables known to \WOPPER/
    and \.{print \textit{prefix}*} causes a listing of all variables
    starting with the given prefix.
  \item \.{debug}$\vert$\.{nodebug} \textit{flag} \hfil\goodbreak
    Toggle debugging flags.
  \item \.{testran} \hfil\goodbreak
    Test the portability of the random number generator.  We use
    a generator of the Marsaglia-Zaman variety \cite{MZT90},
    which should give
    identical results on almost all machines.
  \item \.{banner} \hfil\goodbreak
    Print a string identifying this version of \WOPPER/.
  \item \.{echo} \textit{message} \hfil\goodbreak
    Print \textit{message} on standard output.
  \item \.{lugive} \textit{string} \hfil\goodbreak
    Pass \textit{string} to \texttt{JETSET}'s \texttt{LUGIVE()} routine.
    See~\cite{JETSET} for details.
  \item \.{artune} \textit{string} \hfil\goodbreak
    Pass \textit{string} to \texttt{ARIADNE}'s \texttt{ARTUNE()} routine.
    See~\cite{ARIADNE} for details.
\end{itemize}

Unique abbreviations of the keywords are accepted, i.e.~\verb+'g 1000'+
generates 1000 events.  The tokens are separated by blanks. Blank
lines and lines starting with a \.{\#} are ignored and may be used for
comments.  For portability, only the first 72 characters of each line
are considered.

On UNIX systems \WOPPER/ reads the default startup files \texttt{.wopper}
in the user's home directory and the current directory, if they exist.

%%%%%%%%%%%%%%%%%%%%%%%%%%%%%%%%%%%%%%%%%%%%%%%%%%%%%%%%%%%%%%%%%%%%%

\subsection{Lower Level Interface}
\label{sec:f77-low-level}

The subroutine \.{wopper (code)} has a single integer parameter.
The parameter \.{code} is interpreted as follows:
\begin{itemize}
  \item{} 0: initialize the generator and write an initialization
      record to \hepevt/.
  \item{} 1: generate an event and store it in \hepevt/.  If \WOPPER/
    has not been initialized yet, the necessary initializations are
    performed, but no initialization record is written.
  \item{} 2: perform final calculations and write the results
      to \hepevt/.
\end{itemize}
Figure \ref{ex:f77-wwdcmd} displays excerpts from a simplified
version of \.{wwdcmd} that make the correspondences between the
two levels of the \f77/ interface explicit.

\begin{example}{Event generation loop}{ex:f77-wwdcmd}
\.{* wwdcmd.f}                                                    \\
\>\>\> \.{subroutine wwdcmd (cmdlin)}                             \\
\>\>\> \.{character*(*) cmdlin}                                   \\
\>\>\> \.{...}                                                    \\
\>\>\> \.{else if (cmdlin.eq.'init')}                             \\
\>\>\> \>\> \.{call wopper (0)}        \C initialize \WOPPER/     \\
\>\>\> \>\> \.{call hepawk ('scan')}   \C print
                                            initialization record \\
\>\>\> \.{else if (cmdlin.eq.'generate')}                         \\
\>\>\> \>\> \.{do 10 n = 1, nevent}                               \\
\>\>\> \>\> \>\> \.{call wopper (1)}      \C generate an event    \\
\>\>\> \>\> \>\> \.{call hepawk ('scan')} \C analyze the event    \\
\.{10} \>\>\>\>\> \.{continue}                                    \\
\>\>\> \.{else if (cmdlin.eq.'close')}                            \\
\>\>\> \>\> \.{call wopper (2)}        \C get total cross
                                                    section, etc. \\
\>\>\> \>\> \.{call hepawk ('scan')}   \C finalize analysis       \\
\>\>\> \.{else}                                                   \\
\>\>\> \.{...}                                                    \\
\>\>\> \.{end}
\end{example}

\WOPPER/'s parameters can be accessed on the lower level by the
subroutine \.{wwpsrv (result, action, name, type, ival, rval, dval,
  lval)}. The parameter is specified by its (lowercase) name in the
\.{character*(*)} string \.{name}.  The string \.{action} is either
\verb+'read'+ or \verb+'write'+ corresponding to whether the parameter
is to be inspected or modified. The type of the parameter (\verb+'int'+,
\verb+'real'+, \verb+'dble'+, or \verb+'lgcl'+) is specified in
\.{type}; it is an input parameter for \verb+'read'+ and an output
parameter for \verb+'write'+.  Depending on this type the value is
passed in \.{ival}, \.{rval}, \.{dval}, or \.{lval}, respectively.  The
following error codes will be returned in the string \.{result}:
\verb+' '+: no error, \verb+'enoarg'+: invalid \.{action},
\verb+'enoent'+: no such parameter, \verb+'enoperm'+: permission
denied, and \verb+'enotype'+: invalid \.{type}.

The protection scheme implemented with this parameter handling has
been described in \cite{ADM+92a}.  Its main purpose is to guarantee
consistency of user defined and computed parameters in the generation
phase of the Monte Carlo.

%%%%%%%%%%%%%%%%%%%%%%%%%%%%%%%%%%%%%%%%%%%%%%%%%%%%%%%%%%%%%%%%%%%%%
\subsection{Command Line Interface for UNIX Systems}
\label{sec:cmdline}

The distributed version of the \WOPPER/ main program reads commands
from the standard input channel and passes them directly to
\.{wwdcmd}.  \WOPPER/ writes results to the standard output channel,
user supplied analysis routines may write additional files with
histograms, etc.

\begin{figure}
{\small\begin{verbatim}
usage: $0 [options] [var=val] ...

  var=val                          assign VALue to WOPPER VARiable
                                 
  -e energy, --beam-energy energy: beam ENERGY (= sqrt(s)/2) in GeV
  -s roots, --roots roots:         sqrt(s) (= 2 beam energy) in GeV
  -n num, --events num:            generate NUM events
  -r num, --report num:            write a summary record every NUM events
  -H script, --hepawk script:      use hepawk SCRIPT
                                 
  -q yes|no, --QED yes|no:         apply QED raditiative corrections
  -Q mc, --QCD mc:                 use QCD Monte Carlo MC
  -a mode, --ac mode:              select anomalous couplings MODE
  -f file, --file file:            source FILE with variable settings
  -R rng, --rangen rng:            use Random Number Generator
  -S seed, --seed seed:            random number SEED
                                 
  -k, --keep:                      keep the temporary run directory
  -o file, --output file:          write output to FILE (implies -p FILE)
  -p file, --preserve file:        salvage FILE from temporary run directory
  -x cmd, --sed cmd:               filter hepawk script through sed CoManD
  -t dir, --tmpdir dir:            use DIR as temporary run directory
  -b bin, --binary bin:            use BIN as WOPPER binary
                                 
  -h, --help:                      this message
\end{verbatim}}
\caption{\label{fig:wrun}%
  Command line options of the \texttt{wrun} script for
  \WOPPER/ on UNIX systems.}
\end{figure}

Using no other operating system services than the standard input
channel for input makes the program very portable. At the same time,
it is not the most user friendly approach for repetitive tasks, like
scanning parameter sets.  Therefore we provide a shell script
\.{wrun}, which can be used to control wopper from the UNIX
commandline.

In figure~\ref{fig:wrun} the usage message of \.{wrun} is shown.
Variable can be set on the command line.  The \.{wrun} script will
create an approriate input file for \WOPPER/:
\begin{commands}
  \item \meta{var}\.{=}\meta{value}:
    sets the \WOPPER/ variable \meta{var} to the value \meta{value}.
    Examples: \.{cc11=false}, \.{sin2w=0.23103}, etc.
\end{commands}
Some of these variables can be changed by an option as well:
\begin{commands}
  \item \.{-e} \meta{energy} or \.{--beam-energy} \meta{energy}:
    set the beam energy, this is equivalent to \.{ebeam=}$E_B$.
  \item \.{-n} \meta{num} or \.{--events} \meta{num}:
    set the number of events, this is equivalent to \.{nevent=}$n$.
  \item \.{-r} \meta{num} or \.{--report} \meta{num}:
    write a summary record every everz \meta{num} events,
    this is equivalent to \.{report=}$n$.
  \item \.{-S} \meta{seed} or \.{--seed} \meta{seed}:
    set the seed of the random number generator, this is equivalent to
    \.{rseed=}$n$. 
\end{commands}
For some variables symbolic names or convenience translations are
available:
\begin{commands}
  \item \.{-s} \meta{roots} or \.{--roots} \meta{roots}:
    set the center of mass energy~$\sqrt s$, this is equivalent to
    \.{ebeam=}$\sqrt s/2$.
  \item \.{-q} \.{yes}$|$\.{no} or \.{--QED} \.{yes}$|$\.{no}:
    toggle electromagnetic radiative corrections, \.{--q yes} is
    equivalent to \.{bstyle=1} and \.{-q no} is equivalent to
    \.{bstyle=0}.
  \item \.{-Q} \meta{mc} or \.{--QCD} \meta{mc}:
    select a hadronization Monte Carlo: \meta{mc} can be \.{parton},
    \.{jetset}, \.{herwig} or \.{ariadne}.
  \item \.{-a} \meta{mode} or \.{--ac} \meta{mode}:
    select an anomalous couplings parameterization: \meta{mode} can be
    \.{no} or \.{SM} for no anomalous couplings, \.{Hagiwara},
    \.{kappa}, \.{lambda} for the $\kappa$-$\lambda$ parameterization
    or \.{delta} for the $\delta$-$x$-$y$ parameterization.
  \item \.{-R} \meta{rng} or \.{--rangen} \meta{rng}:
    select a random number generator: \meta{rng} can be \.{ranmar},
    \.{ranlux} or \.{Knuth}.
\end{commands}
The behavious of the script can be changed with the following options:
\begin{commands}
  \item \.{-b} \meta{bin} or \.{--binary} \meta{bin}:
    select a different \WOPPER/ binary.
  \item \.{-o} \meta{file} or \.{--output} \meta{file}:
    write the \WOPPER/ output to \meta{file}
  \item \.{-t} \meta{dir} or \.{--tmpdir} \meta{dir}:
    specify a temporary run directory.  By default, \WOPPER/ will be
    run in a temporary directory with a unique name.
  \item \.{-p} \meta{file} or \.{--preserve} \meta{file}:
    copy the file \meta{file} to the starting directory before
    deleting the temporary run directory.
  \item \.{-k} or \.{--keep}:
    don't delete the temporary run directory.
\end{commands}
The following options are useful if \hepawk/~\cite{Ohl92a} is used for
event analysis:
\begin{commands}
  \item \.{-H} \meta{script} or \.{--hepawk} \meta{script}:
    select a \hepawk/ script.
  \item \.{-x} \meta{cmd} or \.{--sed} \meta{cmd}:
    filter the \hepawk/ script through the UNIX stream editor
    \.{sed}.  This is particularly useful for changing the name of the
    file in which \hepawk/ saves its histograms.  For example, if the
    script contains the command \verb+save("@@@");+, the option
    \.{-x s/@@@/foo.hbook/} will cause \hepawk/ to save the
    histograms in \.{foo.hbook}.
\end{commands}

\begin{figure}
{\small\begin{verbatim}
#!/bin/sh
run () {
  ./wrun -a delta ac$1 -q $2 -e 95.0 -n 100000 cc11=false \
    scheme=-2 ahpla=128.07 sin2w=0.23103 mass1z=91.1888 \
    mass1w=80.23 gamm1z=2.4974 gamm1w=2.08 \
    -o ac.$3.out -p ac.$3.hbook -H ac.hepawk \
    -x s/ac.hbook/ac.$3.hbook/
}
for q in 0 1; do
  run acpara=0 $q sm__.q$q
  for d in 2 5; do
    for v in dz xg xz yg yz zz; do
      for s in - +; do
        run $v=${s}0.$d $q $v$s$d.q$q
      done
    done
  done
done
exit 0
\end{verbatim}}
\caption{\label{fig:acloop}%
  Running tests for various anomalous couplings with
  \WOPPER/ on UNIX systems.}
\end{figure}
This shell script can be used to run test suites easily.  For example,
the tests in~\cite{BvS95} can be run with the shell script in
figure~\ref{fig:acloop} provided that the file \.{ac.hepawk} contains
an appropriate \hepawk/ script.

%%%%%%%%%%%%%%%%%%%%%%%%%%%%%%%%%%%%%%%%%%%%%%%%%%%%%%%%%%%%%%%%%%%%%
\section{Conclusions}
\label{sec:concl}

We have presented version \Version/ of the Monte Carlo event generator
\WOPPER/ for $W$ pair production and decays into four fermions at high
energy $e^+e^-$ colliders.  The distinguishing feature of \WOPPER/ is
the inclusion of higher order electromagnetic corrections including soft
photon exponentiation and explicit generation of exclusive hard photons.
In contrast to fixed order calculations which have to be exponentiated
by hand, \WOPPER/ handles the multiphoton effects explicitly.

The present version does not contain weak corrections.  Forthcoming
versions of the Monte Carlo generator will include weak corrections in
the framework of effective Born cross sections \cite{DBD92:FKK+92}.

%%%%%%%%%%%%%%%%%%%%%%%%%%%%%%%%%%%%%%%%%%%%%%%%%%%%%%%%%%%%%%%%%%%%%
\section*{Acknowledgments}
\label{sec:ack}

Thomas Mannel and Angelika Himmler have contributed to earlier
versions of \WOPPER/.  We are grateful to the members of the
$W$-physics and event generator working groups of the 1995 LEP2
workshop for appreciating our efforts.

%%%%%%%%%%%%%%%%%%%%%%%%%%%%%%%%%%%%%%%%%%%%%%%%%%%%%%%%%%%%%%%%%%%%
\appendix
%%%%%%%%%%%%%%%%%%%%%%%%%%%%%%%%%%%%%%%%%%%%%%%%%%%%%%%%%%%%%%%%%%%%
\section{Distribution}

The latest release of \WOPPER/ is available by anonymous ftp from
\begin{verbatim}
  crunch.ikp.physik.th-darmstadt.de
\end{verbatim}
in the directory
\begin{verbatim}
  pub/ohl/wopper
\end{verbatim}
or on the World Wide Web at the URL
\begin{verbatim}
  http://crunch.ikp.physik.th-darmstadt.de/monte-carlos.html#wopper
\end{verbatim}
Important announcements (new versions, fatal bugs, etc.)
will be made through the mailing list 
\begin{verbatim}
  wopper-announce@crunch.ikp.physik.th-darmstadt.de
\end{verbatim}
Subscriptions can be obtained from
\begin{verbatim}
  majordomo@crunch.ikp.physik.th-darmstadt.de
\end{verbatim}
(send a message consisting of \texttt{help} to \texttt{majordomo} for
instructions on how to subscribe, don't send such messages to the list
itself).

%%%%%%%%%%%%%%%%%%%%%%%%%%%%%%%%%%%%%%%%%%%%%%%%%%%%%%%%%%%%%%%%%%%%%%%%%%%%%%%
\section{Installation}

\WOPPER/ is distributed in PATCHY format \cite{KZ88}.  Plain \f77/
versions can be made available on request.  The source conforms to ANSI
X3.9-1978, with the exception of use of the following rather common
extensions:
\begin{itemize}
  \item \.{implicit none} (may be disabled),
  \item \.{double complex} arithmetic (essential), and
  \item \.{do \ldots{} end do} constructs.
\end{itemize}
Therefore, it should run without modifications on all platforms.

\subsection{UNIX Systems}

On UNIX systems, the configuration, compilation and installation
can be performed automatically according to the following sequence:
\begin{verbatim}
    $ ./configure
    $ make
    $ make install
\end{verbatim}
Optional interface code for external hadronization Monte Carlos,
analysis routines, etc.~can be selected by giving command line options
to the \.{configure} script.  Figure~\ref{fig:configure} shows some of
these command line options.  This \texttt{configure} script has been
created by the popular GNU Autoconf~\cite{autoconf} package and should
work on all UNIX variants.

If the \.{--with-hepawk} option has been specified in the
\.{configure} step, a self test can be performed by \.{make test}.

\begin{figure}
{\small\begin{verbatim}
Usage: configure [options] [host]
Options: [defaults in brackets after descriptions]
Configuration:
  --cache-file=FILE       cache test results in FILE
  --help                  print this message
  --no-create             do not create output files
Host type:
  --build=BUILD           configure for building on BUILD [BUILD=HOST]
  --host=HOST             configure for HOST [guessed]
  --target=TARGET         configure for TARGET [TARGET=HOST]
Features and packages:
  --disable-FEATURE       do not include FEATURE (same as --enable-FEATURE=no)
  --enable-FEATURE[=ARG]  include FEATURE [ARG=yes]
  --with-PACKAGE[=ARG]    use PACKAGE [ARG=yes]
  --without-PACKAGE       do not use PACKAGE (same as --with-PACKAGE=no)
--enable and --with options recognized:
  --with-g77              use GNU Fortran 77
  --enable-verbose-patchy display all patchy output
  --enable-internal       do not use this!
  --enable-notime         do not use timing functions
  --enable-pedantic       no IMPLICIT NONE
  --with-libpath=PATH     use PATH for libraries
  --with-srcpath=PATH     use PATH for source files (CARs)
  --with-hepawk           use HEPAWK for event analysis
  --with-jetset           use JETSET hadronization
  --with-herwig           use HERWIG hadronization
  --with-ariadne          use ARIADNE hadronization
  --with-circe            use CIRCE for beamstrahlung
  --with-ranlux           use RANLUX generator
  --with-cernlib          use CERNLIB
  --with-lepbench         link in the LEP2 benchmarking code
  --enable-debug          compile for debugging
  --enable-paper-a4       use European (A4) paper
  --enable-paper-us       use US (letter) paper
\end{verbatim}}
\caption{\label{fig:configure}%
  Some of the comandline options of the \texttt{configure} script for
  \WOPPER/ on UNIX systems.}
\end{figure}

\subsection{Non-UNIX Systems}

For non-UNIX systems configuration and compilation has to be performed
manually from the \texttt{CARDS} file and the \texttt{cradle}s.

%%%%%%%%%%%%%%%%%%%%%%%%%%%%%%%%%%%%%%%%%%%%%%%%%%%%%%%%%%%%%%%%%%%%
\section{External Symbols:
         Common Blocks and Subroutines}
\label{sec:ext-names}

To avoid possible name clashes with other packages, all external symbols
exported by \WOPPER/ proper begin with the two letters \.{WW}, except
for the routine \.{wopper} itself and the \hepevt/ common block.

%%% UNIX:
%%% $ nm wopper.o | awk ' $3 ~ /^ww/ { print $3}' | sort

\begin{itemize}
  \item Common Blocks: \hfil\goodbreak
    The following common blocks are used by \WOPPER/:
    \begin{itemize}
    \item \.{/hepevt/, /hepspn/}: standard common blocks for passing
      generated events \cite{AKV89}
    \item \.{/wwpcom/}: main parameter common block, holds all physical
        parameters.  Application programs should access this common block
        through the \.{wwpsrv} routine
    \item \.{/wwcbrn/}: internal parameters used for born cross section
    \item \.{/wwcdec/}: internal parameters used for $W$ decays
    \item \.{/wwcevt/}: internal parameters used for event generation
    \item \.{/wwcsta/}: statistics
    \item \.{/wwctri/}: storage for keyword lookup
    \end{itemize}
  \item Driver Program: \hfil\goodbreak
    \begin{itemize}
    \item \.{wwdriv}: sample main program, which reads commands
      from standard input and feeds them into \.{wwdcmd}
    \item \.{wwdloo}: command loop, reading command from a terminal or
      file and executing them
    \item \.{wopper}: the low level entry point into the generator
      for application programs
    \item \.{wwdcmd}: \WOPPER/'s command interpreter, the preferred
      entry point for application programs.  Executes a single command
    \item \.{wwdlxi, wwdlxd, wwdlxs}: Utility routines: tokenization
      of input
    \item \.{wwdsig}: UNIX signal handler
    \item \.{isatty}: check if this job is run interactively
  \end{itemize}
  \item Parameter Management: \hfil\goodbreak
    These routines are used to control the parameters common block
    \.{/wwpcom/}:
  \begin{itemize}
    \item \.{wwpsrv}: server handling parameter changing requests
    \item \.{wwpini}: \.{block data} supplying default values
    \item \.{wwpprn}: print parameters
  \end{itemize}
  \item Initialization:\hfil\goodbreak
  \begin{itemize}
    \item \.{wwinit}: main entry point for initializations
    \item \.{wwigsw}: initialization of electroweak parameters
    \item \.{wwicut}: initialization of internal Monte Carlo parameters
    \item \.{wwibmx}: find maximum of Born cross section
    \item \.{wwibmy}: wrapper for \.{wwuamo}
    \item \.{wwibn}: auxiliary function for finding maximum of on-shell
      cross section
    \item \.{wwibns}: auxiliary function for finding maximum of
      cross section for symmetric virtualities
    \item \.{wwibna}: auxiliary function for finding maximum of
      cross section for one $W$ on-shell
    \item \.{wwibnv}: auxiliary function for finding maximum of
      cross section for general case
  \end{itemize}
  \item Final calculations:
  \begin{itemize}
    \item \.{wwclos}: calculate total cross section, errors and close
      the generator
    \item \.{wwstat}: statistics
  \end{itemize}
  \item Hard Subprocess Generation:\hfil\goodbreak
  \begin{itemize}
    \item \.{wwgen}: main entry point for hard subprocess generation
    \item \.{wwgcfs}: selection of initial and final state quantum
      numbers
    \item \.{wwgww}: generation of angular distribution of pseudo $W$'s
    \item \.{wwgppr}: generate four-momenta of pseudo $W$'s
    \item \.{wwgdec}: generate final state fermions from pseudo $W$
      decays
    \item \.{wwdqfl}: select quark flavors in W decay
  \end{itemize}
  \item Branching: \hfil\goodbreak
  \begin{itemize}
    \item \.{wwbini}: generates the initial state photon radiation
%    \item \.{wwbfin}: generates the final state photon radiation
  \end{itemize}
  \item Cross Sections: \hfil\goodbreak
  \begin{itemize}
    \item \.{wwheli}: coefficients of the helicity amplitude
        decomposition
    \item \.{wwxtot}: total off-shell cross section
    \item \.{wwxint}: auxiliary function for integration over pseudo
      $W$ masses
    \item \.{wwxdif}: differential $W$ cross section
    \item \.{wwxdmx}: maximum estimate of differential $W$ cross section
    \item \.{wwxhel}: helicity amplitude for $e^+e^- \to 4f$
    \item \.{wwxhmx}: maximum estimate of helicity amplitude for $e^+e^-
        \to 4f$
  \end{itemize}
  \item Accessing \hepevt/: \hfil\goodbreak
  \begin{itemize}
    \item \.{wweeni}: enter identification of Monte Carlo and run
    \item \.{wweens}: write summary record to \hepevt/
    \item \.{wweent}: enter one particle into \hepevt/
    \item \.{wwenul}: enter null particle into \hepevt/
    \item \.{wwenew}: clear \hepevt/
    \item \.{wwecpy}: copy a \hepevt/ entry
    \item \.{wwesft}: shift \hepevt/ entries
    \item \.{wwepsh}: push \hepevt/ onto a stack (one level deep)
    \item \.{wwepop}: pop \hepevt/ from a stack
  \end{itemize}
  \item Hadronization: \hfil\goodbreak
  \begin{itemize}
    \item \.{wwpart}: leave partons alone
    \item \.{wwlund}: \texttt{JETSET} interface code
    \item \.{lu4frm}: \texttt{JETSET} interface code
    \item \.{wwhwig}: \texttt{HERWIG} interface code
    \item \.{hw4fdo}, \.{hw4fin}, \.{hw4fgo}, \.{hw4fcc}, \.{hw4fcs}:
      \texttt{HERWIG} interface code
    \item \.{hwaend}: \texttt{HERWIG} abnormal end routine
    \item \.{wwaria}: \texttt{ARIADNE} interface code
  \end{itemize}
  \item Random Numbers: \hfil\goodbreak
  \begin{itemize}
    \item \.{wwrgen}: returns a double precision uniform deviate
    \item \.{wwurng}: (\.{subroutine} version)
    \item \.{wwrmz}: random number generator \.{RANMAR}
    \item \.{wwrtst}: test the portability of the random number
                        generator
    \item \.{wwrtmz}: test \.{RANMAR}
    \item \.{wwrkng}: Knuth's random number generator
    \item \.{wwrknu}: Knuth's random number generator (buffered version)
    \item \.{wwrkns}: seed Knuth's random number generator
    \item \.{wwrknl}: set luxury for Knuth's random number generator
    \item \.{wwrkni}: initialize Knuth's random number generator
    \item \.{wwrknt}: test Knuth's random number generator
  \end{itemize}
  \item Utilities: \hfil\goodbreak
  \begin{itemize}
    \item \.{wwumsg}: messages and error exit
    \item \.{wwulwr}: convert input to lower case
    \item \.{wwuboo}: boost a four vector
    \item \.{wwutim}: still available CPU time for this job
    \item \.{wwuamo}: multidimensional minimization
    \item \.{wwumin}: onedimensional minimization
    \item \.{wwuons}: Gram-Schmidt procedure
    \item \.{wwuort}: another Gram-Schmidt procedure
  \end{itemize}
  \item Canonical cuts: \hfil\goodbreak
  \begin{itemize}
    \item \.{adloth}: apply canonical cuts in \hepevt/
    \item \.{adloip}: inner product of vectors in \hepevt/
    \item \.{adloan}: angle to beam in \hepevt/
    \item \.{adload}: add four momenta in \hepevt/
  \end{itemize}
  \item LEP2 benchmarks~\cite{CERN-96-01}: \hfil\goodbreak
    (optional, selected with \.{configure --with-lepbench})
  \begin{itemize}
    \item \.{hepawk}: main analysis routine
    \item \.{lbfill}: fill a  histogram
    \item \.{lbcheb}: Checbycheff moments
    \item \.{lbpowr}: power moments
    \item \.{lbprin}: print output
    \item \.{lblsct}: select event
  \end{itemize}
  \item Keyword search: \hfil\goodbreak
    (using the dynamic tries described in \cite{Dun91})
  \begin{itemize}
    \item \.{wwtins}: insert a new keyword
    \item \.{wwtlup}: look up a (possibly abbreviated) keyword
    \item \.{wwtnew}: insert new a node into the trie
    \item \.{wwtlen}: calculate length of keyword
    \item \.{wwtc2a}: convert keyword from \.{character*(*)} to
      \.{integer(*)}
  \end{itemize}
\end{itemize}

%%%%%%%%%%%%%%%%%%%%%%%%%%%%%%%%%%%%%%%%%%%%%%%%%%%%%%%%%%%%%%%%%%%%

%%%%%%%%%%%%%%%%%%%%%%%%%%%%%%%%%%%%%%%%%%%%%%%%%%%%%%%%%%%%%%%%%%%
\section*{Test Run}

\WOPPER/ version \Version/ is distributed together with a sample command
file and \hepawk/ script, which are given below.  To run this example,
the user will need to link \WOPPER/ with the CERN library, because
histogramming is done by HBOOK \cite{HBOOK}.

The file \texttt{sample.wopper} is read from standard input (\texttt{unit
  stdin}, which is initialized to 5), and \texttt{sample.hepawk} is read
from the file \texttt{SCRIPT} (i.e.\ under MVS from the file which has been
allocated to the \texttt{DDNAME SCRIPT} and under UNIX from the file
\texttt{script} or from the value of the environment variable \texttt{SCRIPT}).

%%%%%%%%%%%%%%%%%%%%%%%%%%%%%%%%%%%%%%%%%%%%%%%%%%%%%%%%%%%%%%%%%%%%%
\subsection*{\texttt{sample.wopper}}

Here is a simple \WOPPER/ command file, setting up parameters and
generating 10000 events.

{\small
\begin{verbatim}
# sample.wopper -- sample WOPPER command file

# parameters
set ebeam  87.5

# run
init
gen 10000
close
quit
\end{verbatim}
} % end \small

%%%%%%%%%%%%%%%%%%%%%%%%%%%%%%%%%%%%%%%%%%%%%%%%%%%%%%%%%%%%%%%%%%%%%

\subsection*{\texttt{sample.hepawk}}

This is a small \hepawk/~\cite{Ohl92a} script that counts the muons
from the $W$ decays and plots a histogram of their energy
distribution.  The first generated event is dumped to illustrate the
usage of the \hepevt/ common block.

{\small
\begin{verbatim}
# sample.hepawk -- sample HEPAWK analyzer for WOPPER.

BEGIN {
  printf ("\nWelcome to the WOPPER test:\n");
  printf ("***************************\n\n");
  printf ("Monte Carlo Version: %s\n", REV);
  printf ("                Run: %d\n", RUN);
  printf ("               Date: %s\n\n", DATE);

  E_max = 100;
  N_chan = 50;
  h_muon_energy
    = book1 (0, "Muon-energy", N_chan, 0, E_beam);
  nr_muons = 0;               # initialize counter
  dumped_an_event = 0;
}

{
  if (dumped_an_event == 0) {
    dump ("vs");              # Dumping first event
    dumped_an_event++;
  }

  for (@p in LEPTONS)
    if (abs(@p:id) == _pdg_muon) {
      fill (h_muon_energy, @p:p:E);
      nr_muons++;
    }
}

END { # Dump some numbers
  printf ("\nRESULTS:\n********\n\n");
  printf ("Total events:    %d, total cross section: %g pb\n",
          NEVENT, XSECT * 1e9);
  printf ("Number of muons: %d\n\n", nr_muons);
  plot();
  printf ("\ndone.\n");
}

\end{verbatim}
} % end \small

%%%%%%%%%%%%%%%%%%%%%%%%%%%%%%%%%%%%%%%%%%%%%%%%%%%%%%%%%%%%%%%%%%%%%

\newpage

\subsection*{\texttt{sample.output}}

The following output should result from the input files above, up to
small roundoff errors and different \f77/ default output formats.

{\small
\begin{verbatim}
wwdcmd: message: Starting WOPPER, Version 1.05/00, (build 960530/2055)
wwdcmd: message:  ... linked with ARIADNE.
wwdcmd: message:  ... linked with CIRCE.
wwdcmd: message:  ... linked with HEPAWK.
wwdcmd: message:  ... linked with HERWIG.
wwdcmd: message:  ... linked with JETSET.
wwdcmd: message:  ... enabled RANLUX.
hepawk: message: starting HEPAWK, Version 1.6
wwigsw: message: *******************************************************
wwigsw: message: "GF scheme" selected:
wwigsw: message: Using GFERMI and ALPHA as input, calculating SIN2W.
wwigsw: message: Using derived W width.
wwigsw: message: Parameters used in this run:
wwigsw: message: AHPLA  =    128.00000 ( = 1/alpha(2 M_W) )
wwigsw: message: SIN2W  =      0.23098 (effective mixing angle)
wwigsw: message: GFERMI = 0.116639E-04 GeV**(-2)
wwigsw: message: ALPHAS =      0.12000
wwigsw: message: GAMM1W =      2.08780 GeV (S.M. value, used)
wwigsw: message: CKMVUS =      0.21960 CKM Matrix (Cabibbo angle)
wwigsw: message: CKMVCB =      0.04000 CKM Matrix
wwigsw: message: CKMVUB =      0.00320 CKM Matrix
wwigsw: message: Z-e-e couplings:
wwigsw: message:    g_V =     -0.01414
wwigsw: message:    g_A =     -0.18586
wwigsw: message: W-e-nu coupling:
wwigsw: message:      g =      0.23050
wwigsw: message: Z-W-W coupling:
wwigsw: message:  g_ZWW =      0.57171
wwigsw: message: gamma-W-W coupling:
wwigsw: message:  g_gWW =      0.31333
wwigsw: message: *******************************************************
wwigsw: message: Using standard model triple gauge couplings.
wwigsw: message: Using energy-dependent width for W and Z propagators.
wwigsw: message: Including "CC11"-type background diagrams.
wwigsw: message: Including initial state radiation (leading logs).
wwigsw: message: No Coulomb correction.
wwigsw: message: *******************************************************
wwinit: message: Selected parton level events.

Welcome to the WOPPER test:
***************************

Monte Carlo Version: v01.05 (May 30 00:00:00  1996)
                Run: 1035996352
               Date: May 30 20:57:00  1996

========================================================================
Dumping the event record for event #         1
There are   13 entries in this record:
------------------------------------------------------------------------
Entry #   1 is an incoming (HERWIG convention) electron
p: (0.8750E+02; 0.0000E+00, 0.0000E+00, 0.8750E+02),  m: 0.0000E+00
------------------------------------------------------------------------
Entry #   2 is an incoming (HERWIG convention) positron
p: (0.8750E+02; 0.0000E+00, 0.0000E+00, -.8750E+02),  m: 0.0000E+00
------------------------------------------------------------------------
Entry #   3 is the CMS system (HERWIG convention)
p: (0.1750E+03; 0.0000E+00, 0.0000E+00, 0.0000E+00),  m: 0.1750E+03
------------------------------------------------------------------------
Entry #   4 is a null entry.
------------------------------------------------------------------------
Entry #   5 is a null entry.
------------------------------------------------------------------------
Entry #   6 is a null entry.
------------------------------------------------------------------------
Entry #   7 is reserved for model builders.
------------------------------------------------------------------------
Entry #   8 is reserved for model builders.
------------------------------------------------------------------------
Entry #   9 is an existing photon
p: (0.1554E+00; 0.6972E-03, -.1679E-03, -.1554E+00),  m: 0.0000E+00
The mother is the positron                  #   5.
------------------------------------------------------------------------
Entry #  10 is an existing muon-neutrino
p: (0.5820E+02; 0.5040E+02, 0.2417E+02, -.1620E+02),  m: 0.0000E+00
The first mother is the W+                        #   7.
The other mother is the anti-muon                 #  11.
------------------------------------------------------------------------
Entry #  11 is an existing anti-muon
p: (0.2788E+02; -.2410E+02, -.6728E+01, 0.1229E+02),  m: 0.1057E+00
The first mother is the W+                        #   7.
The other mother is the muon-neutrino             #  10.
------------------------------------------------------------------------
Entry #  12 is an existing muon
p: (0.4687E+02; -.2813E+00, -.3362E+02, 0.3265E+02),  m: 0.1057E+00
The first mother is the W-                        #   8.
The other mother is the anti-muon-neutrino        #  13.
------------------------------------------------------------------------
Entry #  13 is an existing anti-muon-neutrino
p: (0.4191E+02; -.2602E+02, 0.1617E+02, -.2859E+02),  m: 0.0000E+00
The first mother is the W-                        #   8.
The other mother is the muon                      #  12.
========================================================================

RESULTS:
********

Total events:         10000, total cross section:  13.06     pb
Number of muons:       2195

1Muon-energy

 HBOOK     ID =         1                                        DATE  30/05/96

      172                                  2
      168                                  X
      164                                  X
      160                               7  X
      156                               X  X
      152                               X5 X5
      148                               XX XX
      144                               XX2XX
      140                             2 XXXXX
      136                            7X7XXXXX
      132                            XXXXXXXX
      128                            XXXXXXXX
      124                            XXXXXXXX7
      120                            XXXXXXXXX
      116                            XXXXXXXXX
      112                        2  XXXXXXXXXX
      108                        X 7XXXXXXXXXX
      104                       5X XXXXXXXXXXX
      100                       XX XXXXXXXXXXX
       96                     7 XXXXXXXXXXXXXX
       92                     X XXXXXXXXXXXXXX
       88                     X XXXXXXXXXXXXXX
       84                     X XXXXXXXXXXXXXXX
       80                     X XXXXXXXXXXXXXXX
       76                     X2XXXXXXXXXXXXXXX
       72                     XXXXXXXXXXXXXXXXX
       68                     XXXXXXXXXXXXXXXXX
       64                     XXXXXXXXXXXXXXXXX
       60                     XXXXXXXXXXXXXXXXX
       56                     XXXXXXXXXXXXXXXXX
       52                     XXXXXXXXXXXXXXXXX
       48                     XXXXXXXXXXXXXXXXX
       44                     XXXXXXXXXXXXXXXXX
       40                     XXXXXXXXXXXXXXXXX
       36                     XXXXXXXXXXXXXXXXX
       32                    XXXXXXXXXXXXXXXXXX5
       28                    XXXXXXXXXXXXXXXXXXX
       24                    XXXXXXXXXXXXXXXXXXX
       20                   2XXXXXXXXXXXXXXXXXXX
       16                   XXXXXXXXXXXXXXXXXXXX
       12                   XXXXXXXXXXXXXXXXXXXX72
        8                   XXXXXXXXXXXXXXXXXXXXXX
        4           25    7XXXXXXXXXXXXXXXXXXXXXXXX7  5

 CHANNELS  10   0        1         2         3         4         5
            1   12345678901234567890123456789012345678901234567890

 CONTENTS 100                   11 11111111111
           10               139700901333554652831
            1.      12    34725329672575901903401943  2

 LOW-EDGE  10        111112222233333444445555566666777778888899999
            1.   2468024680246802468024680246802468024680246802468

 * ENTRIES =       2195      * ALL CHANNELS = 0.2195E+04      * UNDERFLOW = 0.00
 * BIN WID = 0.2000E+01      * MEAN VALUE   = 0.4615E+02      * R . M . S = 0.10

done.
wwdriv: message: bye.
\end{verbatim}
} % end \small

%%%%%%%%%%%%%%%%%%%%%%%%%%%%%%%%%%%%%%%%%%%%%%%%%%%%%%%%%%%%%%%%%%%%

\section{Revision History}
\label{sec:history}

\subsection*{Version 1.5, June 1996}
\begin{itemize}
  \item{} Anomalous couplings.
  \item{} ``CC11'' diagrams.
  \item{} LEP2 standardized \.{ARIADNE}, \.{JETSET} and \.{HERWIG}
    interfaces. 
\end{itemize}

\subsection*{Version 1.4, Fall 1995}
\begin{itemize}
  \item{} LEP2 workshop.
\end{itemize}

\subsection*{Version 1.3, April 1995}
\begin{itemize}
  \item{} Canonical cuts and input parameters.
  \item{} Fixed inconsistent phase conventions, which resulted in
    wrong angular distributions.
\end{itemize}

\subsection*{Version 1.2, July 1994}
\begin{itemize}
  \item{} Coulomb correction.
  \item{} Improved \.{JETSET} and \.{HERWIG} interfaces.
\end{itemize}

\subsection*{Version 1.1, February 1994}
\begin{itemize}
  \item{} Hadronization, \.{JETSET} and \.{HERWIG} interfaces.
  \item{} Minor bug fixes.
\end{itemize}

\subsection*{Version 1.0, 1993}

First official release, submitted to the \textit{Computer Physics
Communication Library}.

\end{document}